\documentclass[a4paper,onecolumn,10pt,accepted=2024-02-01,longbibliography]{quantumarticle}
\pdfoutput=1
\usepackage[utf8]{inputenc}
\usepackage[english]{babel}
\usepackage[T1]{fontenc}
\usepackage{amsmath}
\usepackage{amssymb}
\usepackage{mathtools}
\usepackage{hyperref}
\usepackage{subcaption}

\usepackage{tikz}
\usetikzlibrary{snakes}

\usepackage{braket}
\usepackage{dsfont}

\newcommand{\be}{\begin{equation}}
\newcommand{\ee}{\end{equation}}
\DeclareMathOperator{\diag}{diag}

\usepackage[normalem]{ulem}

\begin{document}

\title{Gravitational quantum switch on a superposition of spherical shells}

\author{Natália S. Móller}
\email{natalia.moller@savba.sk }
\affiliation{RCQI, Institute of Physics, Slovak Academy of Sciences, \\ Dúbravská Cesta 9, 84511 Bratislava, Slovakia}
\orcid{0000-0003-0949-9023}
\author{Bruna Sahdo}
\email{brunasahdo@ufmg.br}
\orcid{0000-0002-5582-5048}
\author{Nelson Yokomizo}
\email{yokomizo@fisica.ufmg.br}
\affiliation{Departamento de F\'isica--ICEx, Universidade Federal de Minas Gerais, \\ CP702, 30161-970, Belo Horizonte, MG, Brazil}
\maketitle

\begin{abstract}
In the absence of a complete theory of quantum gravity, phenomenological models built upon minimal assumptions have been explored for the analysis of possible quantum effects in gravitational systems. Implications of a superposition of geometries have been considered in such models, including the occurrence of processes with indefinite order. In a gravitational quantum switch, in particular, the order of operations applied by two agents on a target system is entangled with the state of the geometry. We consider a model describing the superposition of geometries produced by distinct arrangements of spherical mass shells, and show that a protocol for the implementation of a gravitational quantum switch can be formulated in such a system. The geometries in superposition are identical in an exterior region outside a given radius, and differ within such a radius. The exterior region provides a classical frame from which the superposition of geometries in the interior region can be probed. One of the agents crosses the interior region and becomes entangled with the geometry, which is explored as a resource for the implementation of the quantum switch. Novel features of the protocol include the superposition of nonisometric geometries, the existence of a region with a definite geometry, and the fact that the agent that experiences the superposition of geometries is in free fall, preventing information on the global geometry to be obtained from within its laboratory.
\end{abstract}

\section{Introduction}

The formulation of a complete theory of quantum gravity applicable to physical regimes where both gravitational and quantum effects can become relevant remains a central problem in fundamental physics. In the absence of a complete theory,  a strategy explored in recent works for the analysis of quantum gravitational effects consists of studying concrete physical setups with the aid of phenomenological models built upon minimal assumptions \cite{Bose,Marletto,Mari,Belenchia,tbell}. In this approach, instead of starting from some proposed full theory of quantum gravity, one postulates how basic features of quantum theory and general relativity combine in a proposed setup, typically involving the superposition of weak gravitational fields, and explores physical consequences in such a model. Such a strategy provides a means to investigate conceptual questions which can hopefully guide the development of tools required for the formulation of a fundamental theory of quantum gravity, as well as to delineate experimental avenues along which quantum gravity effects might be observed. Recent predictions of phenomenological models have the potential to rule out the possibility that the gravitational field is purely classical, including the production of entanglement mediated by a gravitational interaction~\cite{Bose,Marletto}, the possibility of decoherence in the recombination of wavepackets of a delocalized massive particle~\cite{Mari,Belenchia} and the superposition of temporal orders~\cite{tbell}.

A basic property often assumed to hold in quantum gravity phenomenology is that the gravitational field $g_{\mu \nu}$ can be prepared in a superposition of classical configurations. For a mass in a superposition of distinct positions, for instance, the gravitational field is assumed to be the superposition of the metrics produced by the mass in each position. A key consequence of the superposition of geometries is that causal relations become indefinite, as they are determined by the metric. In particular, the order between two events $\mathcal{A}$ and $\mathcal{B}$ can become indefinite \cite{tbell}, with nonzero amplitudes for spacetimes in which the event $\mathcal{A}$ is in the past of $\mathcal{B}$, as well as for spacetimes in which $\mathcal{B}$ is in the past of $\mathcal{A}$. The development of a framework for probabilistic theories without a fixed background of causal relations was emphasized in \cite{Hardy} as a step towards the formulation of a full theory of quantum gravity. Generalizations of quantum theory that do not assume a fixed background of causal structure were later introduced in \cite{Chiribella,Oreshkov}, and provide tools for the description of processes with an indefinite order among its events. The question of how such abstract formalisms relate to concrete gravitational scenarios can be approached, for instance, through the construction of protocols implementing processes with indefinite order in phenomenological models \cite{tbell}, and might require the further development of techniques for the description of events and causal relations on quantized spacetimes, which can be based on an operational approach \cite{tbell,Rovelli}.

The simplest example of a process with indefinite order is the quantum switch \cite{Chiribella}. In this process, two operations $\mathcal{A}$ and $\mathcal{B}$ are performed on a target system, and the order of the operations is entangled with a control bit $C$. Let $\ket{0}_C$ and $\ket{1}_C$ be orthogonal states of $C$. If the control bit is in the state $\ket{0}_C$, the operation $\mathcal{A}$ is applied before the operation $\mathcal{B}$. If in the state $\ket{1}_C$, the operations are applied in the opposite order.
Preparing the control bit in a superposition $(\ket{0} + \ket{1})/\sqrt{2}$, the target will evolve into a superposition of states obtained through the application of the operations $\mathcal{A}$ and $\mathcal{B}$ in switched orders,
\begin{equation}	\label{eq:quantum-switch}
\frac{\left( \ket{0}_c + \ket{1}_c \right)}{\sqrt{2}} \ket{\psi} \mapsto \frac{ \ket{0}_c \mathcal{B} \mathcal{A} \ket{\psi} + \ket{1}_c \mathcal{A} \mathcal{B} \ket{\psi} }{\sqrt{2}}\, .
\end{equation}
The order of the operations is then indefinite. If the control bit is the gravitational field, the process is called a gravitational quantum switch.

A protocol for the implementation of a gravitational quantum switch in the gravitational field produced by a mass in a superposition of two positions was proposed in \cite{tbell}. An operational approach is there employed to define what is meant by an event in a superposition of spacetimes. One assumes that, inside a laboratory A, procedures can be specified that correspond to the application of an operation $\mathcal{A}$ on a target system that reaches the laboratory and then leaves it. In addition, it is assumed that the proper time can be measured within the laboratory, so that the procedures can be performed at a chosen proper time. The interaction with a target system at a specific proper time at the laboratory corresponds to an event in the superposition of geometries, at which the operation $\mathcal{A}$ is performed. Similarly, a laboratory B can perform an operation $\mathcal{B}$ on the target system at a specific proper time. The laboratories and target system live in a superposition of two geometries, which in the model considered in \cite{tbell} can be arranged so that in one branch of the geometry the operation $\mathcal{A}$ is performed on the target before the operation $\mathcal{B}$, while in the other branch the operations are performed in the opposite order, leading to an implementation of the quantum switch.

Despite having been introduced for the description of events in a superposition of spacetimes, the operational approach can also be applied for systems on a definite, classical spacetime. Operationally defined events differ from the classical notion of events as coincidence points of classical worldlines in the spacetime manifold, and can be delocalized on the background geometry. It turns out that, for operational events, the quantum switch can also be implemented on a classical geometry, if the laboratories are quantum systems that are delocalized \cite{MSY}. In fact, it was argued in \cite{Foo} that when the geometries in superposition are related by a diffeomorphism, as in the protocol described in \cite{tbell}, one can always re-express such a situation as one in which the laboratories are delocalized on a definite spacetime. If a quantum switch were implemented on a superposition of nonisometric geometries, on the other hand, such a direct translation into an equivalent protocol on a classical spacetime would not be available. Superpositions of nonisometric geometries can also produce other effects as superpositions of vacuum fluctuations of a quantum field, which can be observed by Unruh-de Witt detectors, as analyzed in \cite{Foo2,Foo3}.

The reference~\cite{Oreshkov} introduces the notion of closed laboratories in order to establish the process matrix formalism, a framework constructed to characterize and explore tasks with indefinite causal order. A closed laboratory is there defined as a system that is isolated from the rest of the world, except for interactions with target systems that can enter and leave the laboratory. In the protocol of reference~\cite{tbell}, on the other hand, the agents follow worldlines that are not geodesics, with the consequence that a nonzero local gravity can in principle be measured by them, which is different in each semiclassical branch of the quantum spacetime. Hence, by measuring the weight of an object, not only the agents would acquire information on the global structure they are immersed in, preventing them to be interpreted as closed laboratories, but the quantum spacetime would also decohere and the quantum switch would not be implemented. Not any system can thus be used as an agent in the protocol of reference~\cite{tbell}, but only those which perform operations that are insensitive to the local gravity. The possibility of constructing a protocol that depends just on the geometry of spacetime and not on the nature of the systems A and B is in this way directly related to the utilization of closed laboratories. A protocol involving only closed laboratories would implement the quantum switch as described by a process matrix, and universally, in the sense that the transformation \eqref{eq:quantum-switch} would take place for arbitrary operations $\mathcal{A},\mathcal{B}$.

In this work, we present a new protocol for a gravitational quantum switch. We consider a quantum geometry formed by the superposition of two nonisometric classical geometries. Both geometries are isotropic, and identical in a region exterior to a radius $R_1$. The shared exterior region describes a classical geometry surrounding a region where the gravitational field is in a superposition of configurations produced by two distinct arrangements of masses, distributed into thin shells. We introduce a protocol for a quantum switch on this quantum geometry in which a laboratory freely falls in the superposition of geometries. In this setup, a quantum analogue of the Einstein elevator thought experiment is implemented. Physical systems inside a sufficiently small laboratory in free fall behave in the same way regardless of the external geometry in which the laboratory is travelling. As a result, one cannot determine from within the laboratory whether the geometry is in a superposition state, as physical systems inside it behave in the same way in either branch of the superposition. This allows our protocol to implement a quantum switch for any choice of operations $\mathcal{A}$ and $\mathcal{B}$ performed at the laboratories.

In comparison with the well-known protocol for a gravitational quantum switch introduced in \cite{tbell}, the main novelties of our work are: (i) the agents that perform the operations are closed laboratories, as required for an implementation of the quantum switch as described in the process matrix formalism; (ii) the geometries in superposition are nonisometric, so that the protocol cannot be directly mapped into an analogous protocol on a definite geometry; (iii) the existence of an exterior region with a definite geometry, which provides a classical frame from where the quantum geometry can be probed by studying how quantum systems transform as they are thrown into the quantum region and emerge from it again into the classical region. As the order of the operations in the proposed protocol is controlled by the gravitational field, its eventual implementation would show that a nonclassical gravitational field can indeed produce an indefinite causal structure.

The paper is organized as follows. In Section \ref{sec:geometries}, we introduce the classical geometries involved in our protocol and describe their quantum superposition in an operational approach. A protocol for the implementation of a quantum switch controlled by the gravitational field in such a superposition of geometries is presented in Section \ref{sec:quantum-switch}. The relation between our protocol and previous implementations of the quantum switch is discussed in the same section. We summarize our results in Section \ref{sec:discussion}.

\section{Superposition of spherical shells}
\label{sec:geometries}

The Schwarzschild metric is a vacuum solution of the Einstein equation that describes the geometry of spacetime around a spherically symmetric localized mass $M$ \cite{chandrasekhar}. It can be written as
\begin{equation}	\label{eq:schwarzschild-metric}
ds^2 = -\left (1- \frac{2M}{r}\right) dt^2 + \left (1- \frac{2M}{r} \right)^{-1} dr^2 + r^2 d\Omega^2 \, ,
\end{equation}
where $d\Omega^2=d\theta^2 + \sin^2 \theta d \varphi^2$ is the metric of the unit $2$-sphere. New solutions of the Einstein equation can be constructed by gluing together pieces of Schwarzschild spacetimes of varied masses. The resulting geometry is a new solution when junction conditions are satisfied at the common boundary of the glued regions \cite{israel,poisson}. In general, a thin mass distribution must be present at this surface. When the glued subregions are spherically symmetric, the shared boundary is occupied by a thin spherical mass shell.

We will consider two distinct spacetimes $\mathcal{M}_1$ and $\mathcal{M}_2$, both obtained by gluing an exterior region formed by patches of Schwarzschild spacetimes to an interior region cut from Minkowski spacetime. The spacetimes are built so that the metrics are the same outside a certain radius $r=R_1$. In addition, the interior flat region in each spacetime extends beyond the Schwarzschild radius of the exterior Schwarzschild metric. As a result, the spacetimes $\mathcal{M}_1$ and $\mathcal{M}_2$ do not have event horizons. We apply techniques regularly explored in quantum gravity phenomenology in order to build a quantum geometry describing the superposition of the spacetimes $\mathcal{M}_1$ and $\mathcal{M}_2$. The resulting nonclassical geometry will be explored in the next section for the formulation of a protocol for a gravitational quantum switch. 

In this section, we first summarize the junction conditions and describe the main properties of the spacetimes $\mathcal{M}_1$ and $\mathcal{M}_2$ required for the formulation of the quantum switch, including the behavior of their bounded radial timelike geodesics. Next, we introduce the assumptions adopted for building a quantum superposition of geometries, and discuss their motivation and interpretation. We then describe how, in a superposition of the spacetimes $\mathcal{M}_1$ and $\mathcal{M}_2$, their isometric exterior regions provide a classical frame from which the superposition of geometries for $r<R_1$ can be probed.

\subsection{Junction conditions}

Let $(V^+,g^+)$ and $(V^-,g^-)$ be solutions of the Einstein equation in regions that meet at a common boundary $\Sigma=\partial V^+ = \partial V^-$, where $V^{\pm}$ is a differentiable manifold and $g^\pm$ is a metric on $V^{\pm}$. We wish to construct a new solution $(V,g)$ on the union $V=V^+ \cup V^-$ that reduces to the previous solutions in each subregion. Let $x_\pm^\alpha$ be coordinate systems defined on the regions $V^\pm$, including on their boundaries, and $g^\pm_{\alpha \beta}$ be the metric in these coordinates. Let $y^a$ be coordinates on the three-dimensional shared surface $\Sigma$. Defining the Jacobian matrices
\begin{equation} \label{eq:jacobian-Sigma}
(e_\pm)^\alpha_a = \frac{\partial x^\alpha_\pm}{\partial y^a} \, ,
\end{equation}
the intrinsic metric of the surface $\Sigma$ induced by the metric on each side of it is given by
\[
h^{\pm}_{ab} = (e_\pm)^\alpha_a (e_\pm)^\beta_b \, g^\pm_{\alpha \beta} \, .
\]
We denote the covariant derivative compatible with the metrics $g_{\alpha\beta}^\pm$ by $\nabla^\pm$, and the unit vector normal to the surface pointing toward $V^+$ by $n_\pm^\alpha$.

The first junction condition states that a new solution of the Einstein equation is obtained only if the metric induced in the common boundary is the same on both sides of the surface:
\begin{equation} \label{eq:first-junction-condition}
h^+_{ab} =  h^-_{ab}\, .
\end{equation}
This condition ensures that a system of coordinates $x^\alpha$ defined on both sides of the surface exists in which the components of the metric $g_{\alpha \beta}$ are continuous. The Christoffel symbols are then well defined, and geodesics can cross the boundary between the glued regions.  We introduce the Jacobian matrix
\[
e^\alpha_a = \frac{\partial x^\alpha}{\partial y^a} \, ,
\]
which relates such coordinates to those on the surface $\Sigma$. When the first junction condition is satisfied, we denote the induced metric $h^+_{ab} =  h^-_{ab}$ on the shared boundary simply by $h_{ab}$.

In general, a distribution of energy and momentum must be present at the shared boundary $\Sigma$. This is not necessary, however, when the second junction condition
\be	\label{eq:second-junction-condition}
K^+_{ab} = K^-_{ab}
\ee
is satisfied, where $K^\pm_{ab}$ is the extrinsic curvature at each side of $\Sigma$,
\be	\label{eq:K-pm}
K^\pm_{ab} = (\nabla^\pm_{\beta} n^\pm_\alpha) (e_\pm)^\alpha_a (e_\pm)^\beta_b \, .
\ee
If the condition \eqref{eq:second-junction-condition} is not satisfied, then a thin matter shell must be present at the surface $\Sigma$. Its contribution to the total energy-momentum tensor is given in the coordinates $x^\alpha$ by
\be	\label{eq:mass-shell-energy-mom}
T_{\Sigma}^{\alpha \beta} = \delta(\ell) S^{ab}  e^\alpha_a e^\beta_b \, ,
\ee
where $\ell$ is the geodesic distance to the surface along a geodesic that crosses it orthogonally, and
\be	\label{eq:S-def}
S_{ab} = -\frac{1}{8 \pi} \left( [K_{ab}]- [K] h_{ab} \right) \, ,
\ee
with
\be	\label{eq:[K]}
[K_{ab}] = K^+_{ab} - K^-_{ab}
\ee
representing the discontinuity of the extrinsic curvature at the surface, and $[K]=h^{ab}[K_{ab}]$. The contribution \eqref{eq:mass-shell-energy-mom} must be added to the energy-momentum tensors of the original solutions $g_{\alpha \beta}^{\pm}$. If the glued metrics are vacuum solutions, then the full energy-momentum tensor of the resulting spacetime is given by Eq.~\eqref{eq:mass-shell-energy-mom}.

\subsection{Glued Schwarzschild metrics}

\begin{figure}
\begin{center}
\includegraphics[scale=0.45]{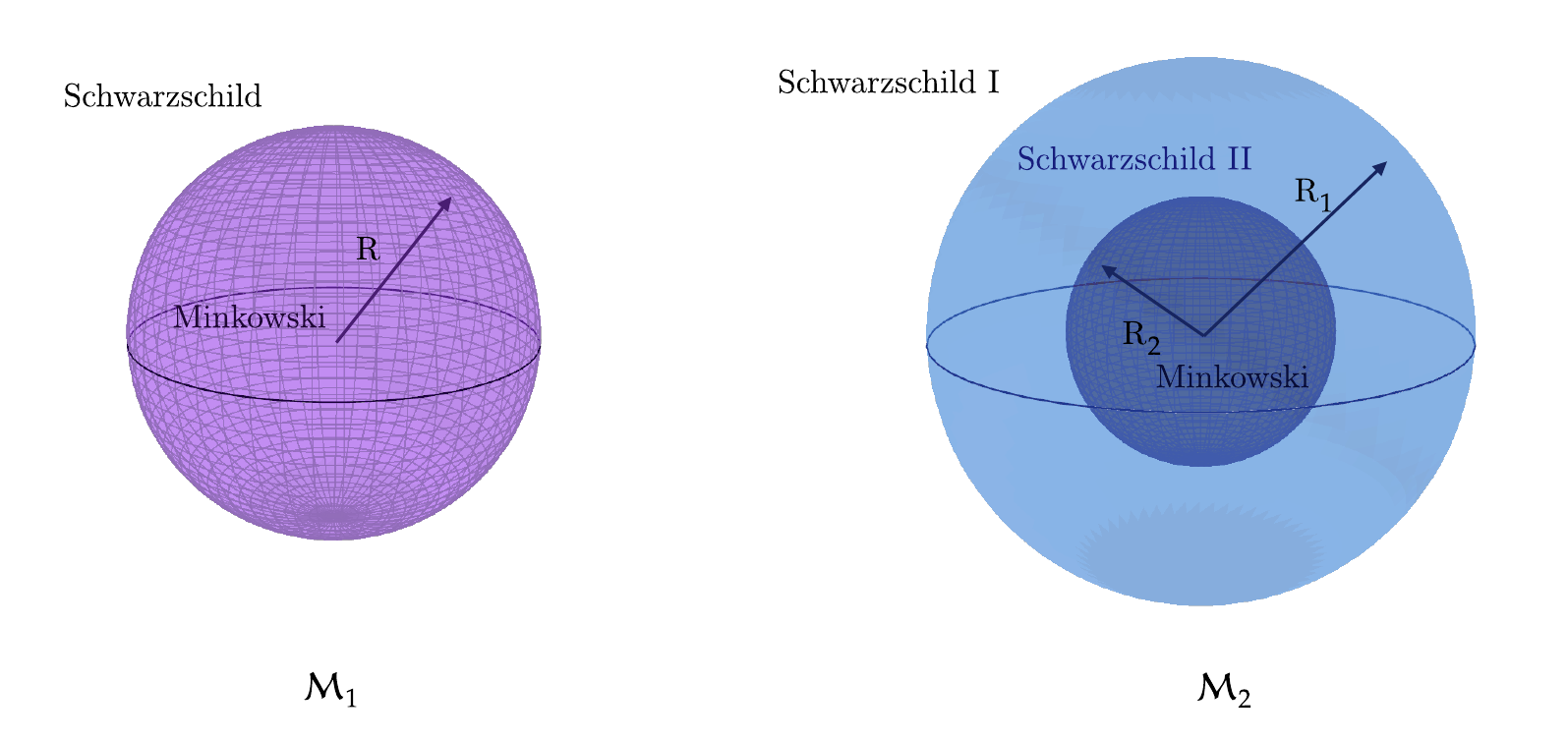}
\end{center}
\caption{Representation of the distinct regions of the spacetimes $\mathcal{M}_1$ and $\mathcal{M}_2$.}
\label{fig:spacetimes}
\end{figure}

\paragraph{Spacetime with one mass shell.} The spacetime $\mathcal{M}_1$ is constructed by gluing the exterior region of a Schwarzschild metric of mass $M$ that lies outside a sphere of radius $R$ to an interior flat region representing the worldvolume of a spatial $3$-ball of the same radius in Minkowski spacetime, as depicted on the left in Figure~\ref{fig:spacetimes}. For the exterior region, we use coordinates
\[
x_+^\alpha = (t_+, r_+, \theta, \varphi) \, , \qquad r_+ \geq R \, ,
\]
and write the Schwarzschild metric in the form \eqref{eq:schwarzschild-metric}. We denote the exterior metric by
\be
ds_+^2 = g^+_{\alpha \beta} dx_+^\alpha dx_+^\beta \,. 
\ee
In the interior region, we use coordinates
\[
x_-^\alpha = (t_-, r_-, \theta, \varphi) \, , \qquad 0 \leq r_- \leq R \, ,
\]
and express the Minkowski metric as
\[
ds_-^2 = g^-_{\alpha \beta} dx_-^\alpha dx_-^\beta = - dt_-^2 + dr_-^2 + r_-^2 d\Omega^2 \, .
\]

The radial coordinates are such that $r_+=r_-=R$ at the interface $\Sigma$ of the two regions. The variables used for the angular coordinates in the interior and exterior regions are identified on $\Sigma$, $\theta_+=\theta_-=:\theta$ and $\varphi_+=\varphi_-=:\varphi$.  The interior and exterior time variables cannot be directly identified, however, if the junction condition \eqref{eq:first-junction-condition} is to be satisfied. We use instead the identification:
\be \label{eq:R-transf-t}
t_+ = \left( 1- \frac{2M}{R} \right)^{-1/2} t_-\, .
\ee
The condition \eqref{eq:first-junction-condition} can then be easily checked. For instance, by putting $t:=t_+$ and choosing coordinates $y^a=(t,\theta,\varphi)$ on $\Sigma$, we find that
\be
h^+_{ab} = h^-_{ab} = \begin{pmatrix}
	-\left(1-\frac{2M}{R}\right) & 0 & 0 \\
	0 & R^2 & 0 \\
	0 & 0 & R^2 \sin^2 \theta
	\end{pmatrix} \, .
\ee

As the intrinsic metric is the same on both sides of the surface, a coordinate system $x^\alpha=(t,\ell,\theta,\varphi)$ exists on a neighborhood of the surface in which the components of the metric vary continuously \cite{israel,poisson}. For instance, one can introduce a new time variable
\be \label{eq:new-t-1-shell}
t = \begin{dcases}
	t_+ \, , & \text{in the exterior region},  \\
	\left( 1- \frac{2M}{R} \right)^{-1/2} t_-\, , & \text{in the interior region},
\end{dcases}
\ee
and a new radial coordinate defined by
\[
\frac{d\ell}{dr_+} = \left(1- \frac{2M}{r_+} \right)^{-1/2} \, , \quad \frac{d\ell}{dr_-} = 1 \, ,
\]
together with the boundary conditions $\ell|_\Sigma = r_+|_\Sigma = r_-|_\Sigma = R$. The metric in the new coordinates $x^\alpha=(t,\ell,\theta,\varphi)$ assumes the form:
\begin{equation}
ds^2=
\begin{dcases}
-\left(1-\frac{2M}{r_+(l)}\right)dt^2+d\ell^2+r_+^2(\ell)d\Omega^2 \, , & \ell \geq R \, ,
\\
-\left(1-\frac{2M}{R}\right)dt^2+d\ell^2+\ell^2d\Omega^2 \, , & \ell< R \, ,
\end{dcases}
\end{equation}
and is explicitly continuous across $\Sigma$.

The continuity of the metric implies that the Christoffel symbols are well behaved, without terms including Dirac deltas. As a result, a geodesic $x^\mu(\tau)$ parametrized by its proper time has a tangent vector $U^\mu = dx^\mu/d\tau$ that varies continuously along the curve. Expressing it at $\Sigma$ in terms of the interior and exterior coordinates, we have:
\be
\left. \frac{dx^\mu}{d\tau} \right|_\Sigma = \left. \left( \frac{\partial x^\mu}{\partial x_\pm^\nu} \frac{dx_\pm^\nu}{d\tau} \right)\right|_\Sigma \, .
\ee
This allow us to determine the relation between the components of the tangent vector on each side of $\Sigma$. We find that:
\begin{align} 
\frac{dr_+}{d\tau}&=\left(1-\frac{2M}{R} \right)^{1/2} \frac{dr_-}{d\tau} \, , \nonumber\\
\frac{dt_+}{d\tau}&=\left( 1-\frac{2M}{R}\right)^{-1/2} \frac{dt_-}{d\tau} \, .
\label{eq:R-transf-U}
\end{align}

The second junction condition \eqref{eq:second-junction-condition} is not satisfied on $\Sigma$. As a resut, the spacetime $\mathcal{M}_1$ must include a mass shell at $r=R$. The calculations of the discontinuity of the extrinsic curvature and of the energy-momentum tensor are straightforward, and presented in Appendix \ref{sec:energy-momentum}.

\paragraph{Spacetime with two mass shells.} The spacetime $\mathcal{M}_2$ is composed of three regions glued along two spherical surfaces, consisting of an exterior region I described by a Schwarzschild metric of mass $M$,
\[
ds_1^2 = -\left(1-\frac{2M}{r_1}\right) dt_1^2 + \left(1-\frac{2M}{r_1}\right)^{-1} dr_1^2 + r_1^2 d\Omega^2 \, ,
\]
with $r_1 > R_1$, an intermediate region II described by a Schwarzschild metric of mass $m$,
\[
ds_2^2 = -\left(1-\frac{2m}{r_2}\right) dt_2^2 + \left(1-\frac{2m}{r_2}\right)^{-1} dr_2^2 + r_2^2 d\Omega^2 \, ,
\]
with $R_2 \leq r_2 \leq R_1$, and an interior flat region III described by the Minkowski metric,
\[
ds_3^2 = - dt_3^2 + dr_3^2 + r_3^2 d\Omega^2 \, ,
\]
with $0 \leq r_3 \leq R_2$, as depicted on the right in Figure~\ref{fig:spacetimes}. We denote the boundary surface between the regions I and II by $\Sigma_1$, and that between the regions II and III by $\Sigma_2$. We set $M>m$.

The proof that the junction condition \eqref{eq:first-junction-condition} is satisfied at $\Sigma_2$ is identical to that for a single shell previously discussed. The time coordinates in the regions II and III are related at the common boundary $\Sigma_2$ through
\be
t_2 = \left(1-\frac{2m}{R_2}\right)^{-1/2} t_3 \, ,
\ee
and the components of the tangent vector to a geodesic curve transform at $\Sigma_2$ according to:
\begin{align}
\frac{dr_2}{d\tau} &=\left(1-\frac{2m}{R_2}\right)^{1/2} \frac{dr_3}{d\tau} \, , \nonumber \\
\frac{dt_2}{d\tau} &=\left( 1-\frac{2m}{R_2}\right)^{-1/2} \frac{dt_3}{d\tau} \, .
\label{eq:R2-transf-U}
\end{align}

The junction condition \eqref{eq:first-junction-condition} is also satisfied at $\Sigma_1$, with the time coordinates in the regions I and II identified at $\Sigma_1$ through
\be	\label{eq:t-cross-sigma-1}
t_1 = \left(1-\frac{2M}{R_1}\right)^{-1/2} \left(1-\frac{2m}{R_1}\right)^{1/2} t_2 \, .
\ee
As in the previous cases, one can construct a system of coordinates in a neighborhood of $\Sigma_1$ in which the components of the metric are continuous (see Appendix \ref{sec:energy-momentum}). The continuity of the components of the tangent vector of a geodesic curve $x^\mu(\tau)$ in such a coordinate system again leads to relations among its components in the original interior and exterior coordinates, which read:
\begin{align} \label{eq:R1-transf-U}
\frac{dr_1}{d\tau}&=\left(1-\frac{2M}{R_1}\right)^{1/2}\left(1-\frac{2m}{R_1}\right)^{-1/2} \frac{dr_2}{d\tau} \, , \nonumber \\
\frac{dt_1}{d\tau}&=\left( 1-\frac{2M}{R_1}\right)^{-1/2}\left( 1-\frac{2m}{R_1}\right)^{1/2} \frac{dt_2}{d\tau} \, ,
\end{align}

The second junction condition is not satisfied at $\Sigma_1$ or $\Sigma_2$. Mass shells must then be present at both surfaces. The discontinuity of the extrinsic curvature and the energy-momentum tensor at these surfaces are computed in Appendix \ref{sec:energy-momentum}.

\subsection{Radial timelike geodesics in glued Schwarzschild metrics}
\label{sec:geodesics}

Let us now discuss bounded timelike radial geodesics on the spacetimes $\mathcal{M}_1$ and $\mathcal{M}_2$ constructed in the previous section. We first review the relevant formulas for radial geodesics in a single Schwarzschild spacetime and then build compositions of such solutions across the glued patches of Schwarzschild and Minkowski spacetimes.

Consider a Schwarzschild metric of mass $M$. Let a massive body be released from rest at the radius $r_i$, at $t=0$. For radial motion, the conserved quantity $E$ associated with invariance under time translations is given by
\be \label{eq:energy-formula}
E^2 = \left(\frac{dr}{d\tau} \right)^2 + \left(1-\frac{2M}{r}\right) \, .
\ee
Computing it at the initial time, we obtain
\be
E= \sqrt{1-\frac{2M}{r_i}} \, .
\ee
Following \cite{chandrasekhar}, we introduce a new radial variable $\eta$ defined by
\be	\label{eq:eta-def}
\eta(r)=2 \arccos \sqrt{\frac{r}{r_i}} \, .
\ee
At the initial position, we have $\eta(r_i)=0$. At the horizon $R_S=2M$, we have
\be
\eta_H := \eta(R_S) = 2 \arcsin E  \, .
\ee

The radial geodesic is described in parametric form in terms of the variable $\eta$ as:
\begin{align}
r(\eta) &= r_i \cos^2 \left( \frac{\eta}{2}\right) \, , 
\label{eq:r-eta} \\
t(\eta) &= E \left( \frac{r_i^3}{2M} \right)^{1/2} \left[ \frac{1}{2} (\eta + \sin \eta)+(1-E^2)\eta \right]+ 2 M \log \left[ \frac{\tan(\eta_H/2)+ \tan(\eta/2)}{\tan (\eta_H/2)-\tan(\eta/2)}\right] \, .
\label{eq:t-eta}
\end{align}
The proper time along the curve is given by
\be \label{eq:tau-eta}
\tau(\eta) = \left( \frac{r_i^3}{8M} \right)^{1/2} (\eta + \sin \eta) \, ,
\ee
and the components of the tangent vector read:
\begin{align} \label{eq:U-parametric}
U^0(\eta) = \frac{dt}{d\tau}(\eta) &= \frac{E \cos^2 (\eta/2)}{\cos^2 (\eta/2)-\cos^2 (\eta_H/2)} \, , \nonumber \\
U^1(\eta) = \frac{dr}{d\tau}(\eta) &= -(1-E^2)^{1/2} \tan(\eta/2) \, .
\end{align}

In both spacetimes $\mathcal{M}_1$ and $\mathcal{M}_2$, the exterior region is a Schwarzschild metric of mass $M$. We wish to determine the radial geodesic followed in each of these spacetimes by a massive particle released from rest at $r_i$. We proceed as follows. The motion is the same in both spacetimes in the common external geometry. When the geodesic reaches a mass shell, we compute the transformation of the components of the tangent vector across the shell. The transformation is given for each of the shells in the considered spacetimes by Eqs.~\eqref{eq:R-transf-U}, \eqref{eq:R2-transf-U} and \eqref{eq:R1-transf-U}. An updated energy is then computed after the shell crossing, and the continuation of the geodesic in the new patch is described again in parametric form using Eqs.~\eqref{eq:r-eta} and \eqref{eq:t-eta}. The process is repeated at each shell crossing. In what follows, we discuss the cases of $\mathcal{M}_1$ and $\mathcal{M}_2$ in detail.

\paragraph{Spacetime with one mass shell.} In the spacetime $\mathcal{M}_1$, we consider a body released from rest at $r_i$ and set $t_+(r_i)=\tau(r_i)=0$. We denote the corresponding geodesic curve by $\gamma_1$. At the radius $R$ of the shell, the variable $\eta$ introduced in Eq.~\eqref{eq:eta-def} assumes the value
\[
\eta_R = 2 \arccos \sqrt{\frac{R}{r_i}} \, .
\]
The coordinate and proper times when it reaches the shell are given by Eqs.~\eqref{eq:t-eta} and \eqref{eq:tau-eta},
\[
\Delta t_+=t_+|_\Sigma = t(\eta_R) \, , \qquad \Delta \tau_+ = \tau|_\Sigma=\tau(\eta_R) \, .
\]
From Eq.~\eqref{eq:U-parametric}, the components of the $4$-velocity are
\[
\left. \frac{dt_+}{d\tau} \right|_\Sigma = U^0(\eta_R) \, , \quad \left. \frac{dr_+}{d\tau} \right|_\Sigma = U^1(\eta_R) \, .
\]
In terms of the interior coordinates, from Eqs.~\eqref{eq:R-transf-t} and \eqref{eq:R-transf-U}, we have
\[
t_-|_\Sigma = \left( 1- \frac{2M}{R} \right)^{1/2} t_+|_\Sigma \, , 
\]
and
\begin{align}
\left. \frac{dt_-}{d\tau} \right|_\Sigma &= \left(1-\frac{2M}{R} \right)^{1/2} U^0(\eta_R) \, , \\
\left. \frac{dr_-}{d\tau} \right|_\Sigma &= \left(1-\frac{2M}{R} \right)^{-1/2} U^1(\eta_R) \, .
\end{align}

After crossing the shell, the particle freely moves in the Minkowski patch and reaches the center of the coordinates after a coordinate time
\begin{equation}
\Delta t_- = - R \left( 1- \frac{2M}{R} \right) \frac{U^0(\eta_R)}{U^1(\eta_R)} \, ,
\end{equation}
during which the amount of elapsed proper time is
\begin{equation}
\Delta \tau_- = - R  \left(1-\frac{2M}{R} \right)^{1/2} \frac{1}{U^1(\eta_R)} \, .
\end{equation}

The particle crosses the shell again, reaches the radius $r_i$ at the opposite side, and starts oscillating. The period of oscillation required for it to return to the initial position is
\[
\Delta t^{(1)} = 4 \left[ \Delta t_+ + \left( 1- \frac{2M}{R} \right)^{-1/2} \Delta t_- \right] \, ,
\]
and the amount of proper time elapsed in one oscillation is
\[
\Delta \tau^{(1)} = 4 (\Delta \tau_+ + \Delta \tau_-) \, .
\]

\paragraph{Spacetime with two mass shells.} In the spacetime $\mathcal{M}_2$, the radius $R_1$ of the exterior shell corresponds to
\[
\eta_{R_1} = 2 \arccos \sqrt{\frac{R_1}{r_i}} \, .
\]
We set $t_1(r_i)=\tau(r_i)=0$, and denote the geodesic curve by $\gamma_2$. The coordinate and proper times when the body reaches the exterior shell $\Sigma_1$ are given by Eqs.~\eqref{eq:t-eta} and \eqref{eq:tau-eta},
\[
\Delta t_1 = t_1|_{\Sigma_1} = t(\eta_{R_1}) \, , \qquad \Delta \tau_1 = \tau|_{\Sigma_1}=\tau(\eta_{R_1}) \, .
\]
From Eq.~\eqref{eq:U-parametric}, the components of the $4$-velocity at $\Sigma_1$ are
\[
\left. \frac{dt_1}{d\tau} \right|_{\Sigma_1} = U^0(\eta_{R_1}) \, , \qquad \left. \frac{dr_1}{d\tau} \right|_{\Sigma_1} = U^1(\eta_{R_1}) \, .
\]
These quantities can be expressed in terms of the coordinates of the region II through Eqs.~\eqref{eq:t-cross-sigma-1} and \eqref{eq:R1-transf-U}, which yield explicit formulas for $t_2|_{\Sigma_1}$, $(dt_2/d\tau)|_{\Sigma_1}$ and $(dr_2/d\tau)|_{\Sigma_1}$.

The geodesic motion in the region II can be described as part of a geodesic in a single Schwarzschild metric of mass $m$. The energy in this \sout{fictitious} space is computed from Eq.~\eqref{eq:energy-formula} as
\[
E_f= \left[ \left( \left. \frac{dr_2}{d\tau}\right|_{\Sigma_1} \right)^2 + \left(1-\frac{2m}{R_1}\right) \right]^{1/2} \, ,
\]
and is associated with the motion of a particle released from rest at the radius
\[
r_{if} = \frac{2m}{1-E_f^2} \,.
\]
The radial variable \eqref{eq:eta-def} at the shells is then given by:
\be
\eta_{f,R_a} = 2 \arccos \sqrt{\frac{R_a}{r_{if}}} \, ,
\ee
with $a=1,2$. At the horizon of the metric with mass $m$, we have
\[
\eta_{f,H}= 2 \arcsin(E_f) \,.
\]

The amounts of coordinate and proper times spent in the region II are given again by Eqs.~\eqref{eq:t-eta} and \eqref{eq:tau-eta}, now applied for the new initial radius and mass,
\begin{align}
\Delta t_2 &= t(\eta_{f,R_2},r_{if},m) - t(\eta_{f,R_1},r_{if},m) \, , \\
\Delta \tau_2 &= \tau(\eta_{f,R_2},r_{if},m) - \tau(\eta_{f,R_1},r_{if},m) \, ,
\end{align}
and the particle reaches the inner shell $\Sigma_2$ at the coordinate time
\be
t_2|_{\Sigma_2} = t_2|_{\Sigma_1} + \Delta t_2 \, .
\ee
From Eq.~\eqref{eq:U-parametric}, the components of the $4$-velocity at the inner shell read:
\begin{align}
\left. \frac{dt_2}{d\tau}\right|_{\Sigma_2} &= \frac{E_f \cos^2 (\eta_{f,R_2}/2)}{\cos^2 (\eta_{f,R_2}/2)-\cos^2 (\eta_{f,H})/2} \, , \nonumber \\
\left. \frac{dr_2}{d\tau}\right|_{\Sigma_2} &= -(1-E_f^2)^{1/2} \tan(\eta_{f,R_2}/2) \, .
\end{align}

Using Eq.~\eqref{eq:R1-transf-U}, the components of the $4$-velocity can be transformed to the interior coordinates, yielding the quantities $(dt_3/d\tau)|_{\Sigma_2}$ and $(dr_3/d\tau)|_{\Sigma_2}$. The coordinate and proper time elapsed until the particle reaches the origin of the coordinates are then easily computed for the uniform motion in the flat interior region,
\begin{align}
\Delta t_3 &= -R_2 \left. \frac{dt_3}{d\tau}\right|_{\Sigma_2} \left( \left. \frac{dr_3}{d\tau}\right|_{\Sigma_2} \right)^{-1} \, , \\
\Delta \tau_3 &= -R_2 \left( \left. \frac{dr_3}{d\tau}\right|_{\Sigma_2} \right)^{-1} \, .
\end{align}

Repeating the procedure of transforming the coordinates and the components of the $4$-velocity at each shell crossing, one finds that the particle again oscillates radially around the origin. Adding the times spent in each patch of the geometry, it is straightforward to compute the period of oscillation taken for the particle to return to its initial position,
\begin{multline}\label{eq:delta-t-2}
\Delta t^{(2)} = 4 \left[ \Delta t_1 + \left(1-\frac{2M}{R_1}\right)^{-1/2}\left(1-\frac{2m}{R_1}\right)^{1/2} \Delta t_2 \right. \\
\left. + \left(1-\frac{2M}{R_1}\right)^{-1/2}\left(1-\frac{2m}{R_1}\right)^{1/2} \left( 1-\frac{2m}{R_2}\right)^{-1/2} \Delta t_3 \right] \, ,
\end{multline}
and the total amount of proper time elapsed in one oscillation,
\be	\label{eq:delta-tau-2}
\Delta \tau^{(2)} = 4 (\Delta \tau_1 + \Delta \tau_2 + \Delta \tau_3) \, .
\ee

\subsection{Assumptions on quantum gravity phenomenology}
\label{sec:assumptions}

In order to build a model for a quantum spacetime describing a superposition of semiclassical states representing the spacetimes $\mathcal{M}_1$ and $\mathcal{M}_2$, we adopt basic assumptions regularly explored in models of quantum gravity phenomenology \cite{Bose,Marletto,Mari,Belenchia,tbell}. It is not our purpose to propose a fresh set of hypotheses for quantum gravity phenomenology, but instead to apply procedures already explored in the literature to a new setup, on which we will formulate a new protocol for a gravitational quantum switch. Later on, we discuss the novel features of our protocol and compare it with previous protocols for the implementation of a quantum switch. Following \cite{tbell}, we adopt an operational approach for the description of events in a nonclassical spacetime. In addition, we assume that the principle of superposition holds for gravity. In this section, we explicitly state the assumptions underlying our model, and discuss their motivation and interpretation. 

Since we are interested in discussing causal relations on a nonclassical spacetime, let us first recall how causal relations are described on a classical spacetime \cite{wald}. Let $E_A$ and $E_B$ be events on a classical spacetime $\mathcal{M}$. To determine the causal relation between these events, it is necessary to check whether a signal can be sent from one event to the other. If a signal can be sent from $E_B$ to $E_A$, then the event $E_A$ is in the future of $E_B$. This occurs whenever there is a causal curve\footnote{A curve is causal when its tangent vector is nonspacelike at all its points.} directed to the future starting at $E_B$ and ending at $E_A$ \cite{wald}. The event $E_A$ is said to be in the past of $E_B$ when $E_B$ is in its future, and the events are spatially related if neither is in the future of the other. As the character of a curve is encoded in the metric tensor, the existence or not of a causal curve connecting two given events is determined by the metric. In this way, the full network of causal relations among events in a classical spacetime is encoded in the metric tensor.

Suppose now that the geometry can display quantum fluctuations, so that it cannot be described by a single definite metric tensor. In this case, the fixed background of causal relations encoded in a definite metric is lost. For instance, a causal curve connecting two events may exist for some configurations of the metric representing fluctuations of the geometry but not for others. In order to analyze what replaces the causal structure of a classical spacetime in a specific quantum gravity scenario, three main points must be addressed. First, a concrete representation of a nonclassical spacetime must naturally be introduced, i.e., a model must be adopted for the quantum geometry. In a nonclassical geometry, it may not be immediately clear how to represent physical events, as they will not lie on a definite spacetime. One must then also specify how events should be represented in the adopted model for the quantum geometry. Moreover, one must be able to determine whether a signal can be exchanged between two events in the superposition of geometries. This requires describing how physical systems---which can be a laboratory, a massive particle or a light ray, for instance---, evolve in the quantum geometry. In the context of a specific model, prescriptions can be introduced to address these questions, without the need of embeddeding them in the context of a full theory of quantum gravity. It is enough to state such prescriptions in a sufficiently clear form for the construction of the model of interest, without the ambition of setting down a formal basis for some general axiomatic system for quantum gravity.

As in other works in quantum gravity phenomenology, we adopt the basic hypothesis that the principle of superposition remains valid in a gravitational context. Configurations of the gravitational field are then described by vector states that can be added together to build quantum superpositions. If a classical spacetime $\mathcal{M}$ is produced by a certain distribution of matter, such a geometry must correspond to a semiclassical state of the quantum geometry, i.e., a quantum state peaked in the given classical geometry with small fluctuations around it, which we represent by $\ket{\mathcal{M}}$. If a distribution of matter is prepared in a superposition of two classical configurations, we assume that the gravitational field is described by the superposition of the corresponding semiclassical states of the geometry. In our protocol, the geometries in superposition will consist of the spacetimes $\mathcal{M}_1$ and $\mathcal{M}_2$ discussed in the previous sections, associated with distinct configurations of spherical mass shells.

In order to describe events in a superposition of geometries, we adopt an operational approach, following \cite{tbell}. In an operational approach, physical events are specified by concrete procedures performed in a laboratory by an experimenter. Consider, for instance, a laboratory equipped with an internal clock. Then the observation of a particular time $\tau$ in the laboratory specifies a physical event. If another observation is performed when the clock displays such a time, this also constitutes a physical event at the same time. The observer can also switch an apparatus on or bring two systems together so that they can interact at a given time. These are examples of operationally defined events associated with an instant of time $\tau$ as measured in the laboratory. In our proposed protocol for a gravitational quantum switch, the relevant events will be operationally defined physical events in a superposition of the spacetimes $\mathcal{M}_1$ and $\mathcal{M}_2$.

On a classical spacetime, the observation of an instant of time $\tau$ in a laboratory $L$ defines a spacetime region associated with the event $E$. Neglecting the dimensions of the region, such an event can be identified with a point in spacetime. Consider now the case of a superposition of spacetimes $\mathcal{M}_i$. One might, for instance, consider a situation in which a second laboratory $L_m$ prepares a distribution of matter in a superposition of distinct semiclassical states $\ket{\psi_i}$. The laboratory $L$ will then live in a geometry described by the superposition of the spacetimes $\mathcal{M}_i$ associated with the distinct states of the matter. We restrict to situations in which the matter distributions in superposition do not produce any singularity\footnote{In the presence of a singularity, a physical system can fall into the singularity and have a maximum value for its proper time. We restrict to situations where the proper time of any physical system can have arbitrary values.}. How should the event $E$ describing the observation of a time $\tau$ in the laboratory $L$ be represented in this quantum geometry? We first note that the observation of the time $\tau$ is consistent with all geometries in superposition. If the matter is observed to be in a state $\ket{\psi_i}$, then the event $E$ will simply correspond to an operational event $E_i$ in the spacetime $\mathcal{M}_i$. Hence, the observation of the event $E$ does not collapse the state of geometry into a definite semiclassical state. Accordingly, we represent the event $E$ by a family of images $E_i$ in the spacetimes in superposition.

Such a representation of a physical event in a superposition of geometries was previously employed in \cite{tbell} for a setup involving a mass prepared in a superposition of two positions and two laboratories that experience the quantized gravitational field produced by the mass in superposition. The proper times of the laboratories are represented by their images in each of the two spacetimes in superposition. Such a representation of a physical event in terms of images in each branch of a superposition of geometries was further discussed in \cite{Giacomini1,Giacomini2}. We adopt the same operational definition of events for our model.

Consider now that a light ray is emitted from a laboratory at an instant of time $\tau$ as measured by a clock in the laboratory. This corresponds to a physical event $E$ that, in a superposition of geometries, is represented by events $E_i$ in the geometries in superposition. If the state of the geometry is observed---for instance, through a measurement of the masses that produced it---, then the light ray will propagate on a semiclasical state $\ket{M_i}$. In this case, the behaviour of the light ray must reduce to that described in classical gravity, i.e., it must follow a null geodesic starting at $E_i$ on $M_i$. Suppose now that the state of the geometry was not measured, so that the light ray propagates on a superposition of semiclassical states of the geometry. In this more general case, we assume that the propagation of the light ray is described by its classical evolution in each branch of the superposition, i.e., that its evolution is controlled by the state of the geometry. For a trivial superposition of a single semiclassical state, we then recover the correct classical limit. The same is assumed for the evolution of massive systems and in the case of internal degrees of freedom, as well as for quantum systems on a superposition of geometries. For instance, a massive body in free fall in a superposition of geometries will be represented by a timelike geodesic in each branch of the superposition.

The hypothesis that a physical system evolves in each semiclassical spacetime in a superposition as described by classical gravity is implicitly adopted in several models of quantum gravity phenomenology, and ensures that a correct classical limit is obtained for the dynamics. The evolution of a physical system must reduce to that described in classical gravity, up to small quantum corrections, in a semiclassical state of the geometry. In \cite{tbell}, for instance, time dilations and light ray propagation are determined using classical equations in each of the considered semiclassical geometries in superposition. In the effect of gravitationally mediated entanglement \cite{Bose,Marletto}, the relevant effect is a superposition of time dilations computed using general relativity, as discussed in \cite{Christodoulou}. 

Let us summarize our assumptions and further comment on their interpretation and relation to the discussed motivations.

\begin{enumerate}
\item The gravitational field can exist in a superposition of semiclassical states $\ket{\mathcal{M}_i}$ associated with classical spacetimes $\mathcal{M}_i$ equipped with metrics $g_i$, where each $g_i$ is a classical solution of the Einstein equation for some distribution of matter $T^{\mu \nu}_i$.
\item An operationally defined event $E$ that does not include an observation of the state of the geometry is represented by images $E_i$ in each spacetime $\mathcal{M}_i$ in superposition, where each $E_i$ is an operationally defined event in the spacetime $\mathcal{M}_i$.
\item Let S be a system whose dynamics in each of the classical spacetimes $\mathcal{M}_i$ is known. The dynamics of the system in a superposition of spacetimes is described by solutions of its evolution equation in each of the spacetimes in superposition.
\end{enumerate}

The assumption 1 is the basic hypothesis that the gravitational field can be prepared in a superposition of semiclassical configurations. If a mass distribution is in a superposition of two distinct configurations, we assume that the gravitational field produced by such masses is a superposition of the geometries for each configuration.

The assumption 2 introduces the representation of an operationally defined event in a superposition of geometries. Under this assumption, a laboratory on a superposition of geometries is represented by a copy of itself in each of the geometries: any proper time $\tau$ at the laboratory is associated with an operational event $E_i$ on each spacetime in the superposition, corresponding to the observation of such a proper time in that spacetime. The assumption 2 characterizes the kinematical setting in a superposition of geometries. The identification of spacetime points in the distinct geometries in superposition can also provide an operational meaning to the superposition of geometries itself, as we will soon discuss.

The assumption 3 allows one to model the evolution of a quantum system on a superposition of geometries from its known dynamics on definite spacetimes. A given system can be found in any of the spacetimes in superposition. In each of these geometries, it simply evolves as dictated by quantum mechanics on such a definite geometry. Introducing a $(3+1)$ foliation with a time coordinate $t_i$ for each spacetime $g_i$, the evolution of a quantum system of interest can then be described by states $\ket{\psi_i,t_i}$ that are solutions of its evolution equation in each spacetime. The joint state of the system and the geometry has the form $\sum c_i \ket{\mathcal{M}_i} \otimes \ket{\psi_i,t_i}$. For a localized object, i.e., a system for which the states $\ket{\psi_i,t_i}$ are well localized around a classical trajectory in each spacetime, a proper time can be assigned along the trajectories. The evolution of the system in each branch of the geometry can then alternatively be presented in terms of the proper time, $\ket{\psi_i,\tau_i}$.

\subsection{Superposition of spherical shells}
\label{sec:superposition-shells}

Let us now describe how the superposition of geometries is operationally defined in our specific setup of interest. In particular, we wish to discuss how the assumption 2 provides an operational meaning to the superposition of spacetimes itself. For this purpose, consider a superposition of semiclassical states of the geometry of the form:
\be	\label{eq:q-metric}
\ket{\Psi}=\frac{1}{\sqrt{2}}(\ket{\mathcal{M}_1}+\ket{\mathcal{M}_2}) \, ,
\ee
where $\mathcal{M}_1$ and $\mathcal{M}_2$ are the spacetimes described in Section \ref{sec:geometries}. Denote their respective metrics by $g_1$ and $g_2$. Let $U_a \subset \mathcal{M}_a$, $a=1,2$, be the exterior region $r \geq R_1$ in each geometry. Such regions are isometric, i.e., the spacetimes are identical in their exterior regions, $g_1|_{U_1}=g_2|_{U_2}\equiv g_{ext}$. It is natural to ask whether these regions can be interpreted as a shared classical region of spacetime, with a definite geometry $g_{ext}$. In order to answer this question, we first note that this must be the case if classical procedures for measuring the exterior geometry $g_{ext}$ can be implemented in the quantum geometry \eqref{eq:q-metric}. This corresponds to the existence of local observers that can communicate with each other through the exchange of light rays, such that the relations between the times at which light rays are emitted by a local observer $O_A$ and later observed by another local observer $O_B$ are well described by the trajectories of light signals in the metric $g_{ext}$.

For instance, consider a pair of local observers $O_A$ and $O_B$ specified in the classical metric $g_{ext}$ by fixed spatial coordinates $\vec x_a = (r_a,\theta,\varphi)$ and $\vec x_b = (r_b,\theta,\varphi)$, where we choose the same angular variables for both observers for simplicity. Put $r_b>r_a$. A light ray emitted by $O_A$ at $t_a$ in the direction of $O_B$ will reach the latter at the coordinate time
\[
t_b = t_a + \int_{r_a}^{r_b} \left( 1-\frac{2M}{r} \right)^{-1} dr \, .
\]
The proper time of an observer static with respect to the spherical coordinates $(t,r,\theta,\varphi)$ is related to the coordinate time at each point by $\tau=\sqrt{|g_{00}(\vec x)|}t$. Therefore, the relation between the proper time $\tau_A$ of $O_A$ when the ray is emitted and the proper time $\tau_B$ of $O_B$ when it is absorbed is
\be
\tau_b= \left(1-\frac{2M}{r_b} \right)^{1/2} \left[ \left(1-\frac{2M}{r_a} \right)^{-1/2} \tau_a + \int_{r_a}^{r_b} \left( 1-\frac{2M}{r} \right)^{-1} dr  \right] \, .
\ee
Analogous relations exist for generic pairs of observers at any fixed spatial coordinates. Now suppose that there are laboratories $L_{\vec x}$ in the superposition state $\ket{\Psi}$, labeled by positions $\vec x$, for which such relations among proper times of emissions and absorptions of light rays in the exterior region are satisfied. Each emission and absorption of a light ray corresponds to an operational event E associated with a definite proper time at a laboratory. From assumption 2, each such event must correspond to events $E_1$ and $E_2$ in the geometries $g_1$ and $g_2$. If these images are identified by the same coordinates $(t(\tau,\vec x),\vec x)$ in the exterior region of both spacetimes, we say that $g_{ext}$ describes a classical patch of the quantum geometry $\ket{\Psi}$. In this case, the spacetime points in the exterior geometry $g_{ext}$ can be interpreted as operational events at the laboratories $L_{\vec x}$, as the communication among the laboratories through the exchange of light signals cannot be distinguished from what would be observed in the classical metric $g_{ext}$.

\begin{figure}
\begin{center}
\includegraphics{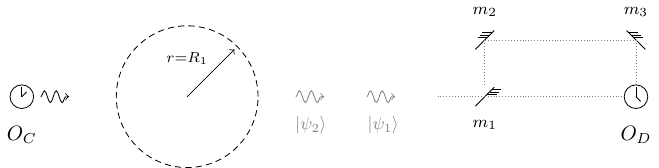}
\caption{A light ray is sent from a local observer $O_C$ to another local observer $O_D$ through the interior region with $r<R_1$. After crossing the interior region, the ray follows a superposition of paths in the exterior region, represented by the wavepackets $\ket{\psi_1}$ and $\ket{\psi_2}$. An arrangement of mirrors $m_i$ ensures that both wavepackets arrive at $O_D$ at the same time. The mirror $m_1$ is removed after reflecting the wavepacket $\ket{\psi_1}$, before it is reached by the wavepacket $\ket{\psi_2}$.}
\label{fig:lightray}
\end{center}
\end{figure}

A local laboratory in the exterior region can also send a light ray toward the interior region $r<R_1$. In this case, the ray will follow distinct geodesics in each of the geometries in superposition. After leaving the interior region, it will carry information on the superposition of geometries. Consider, for instance, that the light ray is emitted from a laboratory $O_C$ at the position $\vec x_c = (r_c,\pi/2,0)$, in the exterior region, along the radial direction toward the interior region. Another local observer $O_D$ situated at $\vec x_d = (r_d,\pi/2,\pi)$, in the exterior region at the other side of the shells, will receive the signal. The emitted light ray crosses the interior region and emerges in the opposite side of the shells. Let $r_i(t)$ describe the geodesic motion of the ray in this region for each spacetime $\mathcal{M}_i$. As the coordinate times taken to cross the interior region in each spacetime $\mathcal{M}_1$ and $\mathcal{M}_2$ are in general different, the ray emerges in the exterior region in a superposition of two localized states, entangled with the geometry (see Fig.~\ref{fig:lightray}),
\be	\label{eq:ray-superposition}
\ket{\Psi} \ket{r,t_c} \mapsto \frac{\ket{\mathcal{M}_1}\ket{\psi_1,t}+\ket{\mathcal{M}_2}\ket{\psi_2,t}}{\sqrt{2}} \, ,
\ee
where $t_c$ is the emission time, $t>t_c$ is large enough so that the ray is in the exterior region for both spacetimes, and the states $\ket{\psi_i,t}$ are wavepackets peaked at $r_i(t)$. As a result, the arrival of the light ray at the laboratory $O_D$ occurs after distinct delays $\Delta t_D^1,\Delta t_D^2$ in each branch of the superposition. Repeating the experiment, the delay $\Delta t_D^1$ will be observed in half the observations, with the same probability for $\Delta t_D^2$. The observation of the time of arrival of the light ray at $O_D$ amounts to a measurement of the geometry of the interior region.

Suppose now that a different experiment is prepared, in which mirrors are arranged in the exterior region at the opposite side of the shells so that the light ray reaches $O_D$ at the same coordinate time in both geometries (see Fig.~\ref{fig:lightray}). For concreteness, consider a configuration of the shells for which $r_1>r_2$ after the ray has crossed the interior region. A mirror can then be placed in a point of the trajectory $r_1(t)$ at a radius $r_m$, with $R_1<r_m<r$, and deviate the trajectory from its original radial direction toward $O_D$, after which the mirror is rapidly removed, so that it does not affect the trajectory $r_2(t)$. Other mirrors can be arranged so that the reflected wavepacket reaches the laboratory $O_D$ at the same time $t_d$ as the second trajectory $r_2(t)$. Under these conditions, the arrival of the ray at $O_D$ will now always be observed after the same coordinate time delay $\Delta t_D$. In addition, the detection of the ray can be made in a manner that, under a postselection, the geometry remains in its original superposition state \eqref{eq:q-metric}. The states $\ket{\psi_i,t_d}$, although peaked at the same position, are in general not identical, as they include distinct phases accumulated along their paths from emission to absorption, and possibly distinct momenta, depending on the arrangement of mirrors. The local observer can then measure the ray in a diagonal basis $\ket{\pm}=(\ket{\psi_1,t_d} \pm \ket{\psi_2,t_d})/\sqrt{2}$, and select runs of the experiment with the result $\ket{+}$. This leaves the geometry in the state \eqref{eq:q-metric}. Such an exchange of a light ray between $O_C$ and $O_D$, tuned by a precise positioning of mirrors dependent on the geometries in superposition, is the  analogue, for a ray crossing the interior region, of the direct exchange of light signals through the exterior region that allows the observers to measure the exterior geometry.

\section{Implementation of the quantum switch}
\label{sec:quantum-switch}

\subsection{Gravitational quantum switch on superposition of mass shells}

Our protocol for the implementation of a gravitational quantum switch is formulated on the quantum geometry given by the superposition of semiclassical states associated with the spacetimes $\mathcal{M}_1$ and $\mathcal{M}_2$ with equal amplitudes, as described by Eq.~\eqref{eq:q-metric}. The exterior regions of both spacetimes describe a classical patch $g_{ext}$ of the geometry, as discussed in the previous section. We use coordinates $(t,\vec x)$ in this region.

\begin{figure}
\begin{center}
\includegraphics{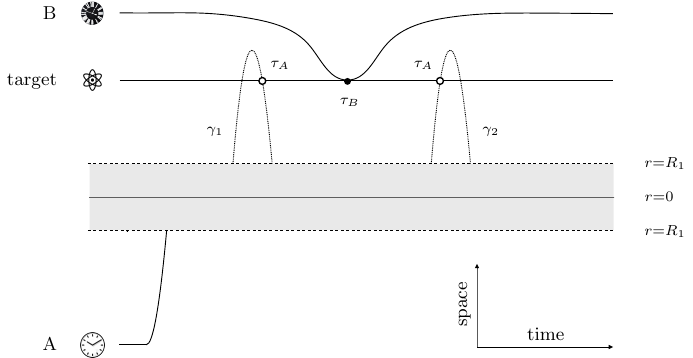}
\end{center}
\caption{Trajectories of the agents A and B and of the target system. The agent A is relased from rest in the exterior region ($r>R_1$) and freely falls toward the shells situated in the interior region ($r<R_1$, represented in gray). After crossing the interior region, it follows distinct geodesics $\gamma_1$ and $\gamma_2$ at the opposite side of the shells for each spacetime in superposition. Each geodesic crosses the worldline of the target, which remains at a fixed position, at the same proper time $\tau_A$. The agent B is brought into contact with the target at its proper time $\tau_B$. The operations are applied at the operational events defined by the proper times $\tau_A$ and $\tau_B$. Afterwards, the agent B is brought back to its initial position, and the geodesics of A oscillate radially until they rejoin at the initial position of the agent. Only the first crossing of the interior region by the agent A is represented.}
\label{fig:trajectories}
\end{figure}

Let us first describe the overall structure of the protocol. Two agents A and B are considered, which perform operations $\mathcal{A}$ and $\mathcal{B}$ on a target system T. The paths of the agents and target are specified in reference to the classical part $g_{ext}$ of the geometry, exterior to the radius $R_1$. For the agent A, an initial position is chosen in this classical region, from which it freely falls toward the interior region in geodesic motion. The agent B moves in a prescribed way in the exterior region, at the opposite side of the shells, and the target system remains at a fixed nearby position. Their paths cross once in the exterior region, when the agent B applies its operation. The agent A meets the target at the same proper time in both branches of the superposition of geometries, and applies its operation. The parameters of the geometries and paths can be chosen so that in one branch of the geometry the target is first acted upon by agent A and then by B, while in the other branch the operations occur in the opposite order, leading to the implementation of a quantum switch. The relevant part of the worldlines of the agents and target system are schematically represented in Figure \ref{fig:trajectories}. We will now describe the trajectories in more detail and show how the quantum switch is implemented.

The agent A is released from rest at an initial radius $r_i> R_1$ in the exterior region, at $t_0=0$. It freely falls toward the center and starts oscillating in the radial direction in both branches of the geometry, following the geodesics $\gamma_1$ and $\gamma_2$ described in Section \ref{sec:geodesics}. As the coordinate times required for the geodesics to cross the interior region $r<R_1$ are different in the classical spacetimes $\mathcal{M}_1$ and $\mathcal{M}_2$, after first crossing the interior region, the agent emerges in the exterior region in a superposition of two paths. The initial radius $r_i$, and the masses and radii of the shells are chosen so that after a number of cycles, the two paths join again and return to the initial position at the coordinate time $t_f$ with the same proper time. The paths of A satisfy the following additional property. Let $r_t<r_i$ be a radius crossed by the agent A in the exterior region. In the first oscillation, at the side opposite to that of the initial position, we denote by $t_A^1$ the coordinate time at which the geodesic in $\mathcal{M}_1$ crosses $r_t$ while travelling back toward the center, and by $t_A^2$ the coordinate time at which the geodesic in $\mathcal{M}_2$ crosses $r_t$ while travelling toward the turning point $r_i$. We require the proper time $\tau_A$ of A to be the same in both cases, with $t_A^1<t_A^2$.

The target remains at a fixed position $\vec x_T$ at the side opposite to that from which the agent A was released, at the radius $r_t$ in the path of A. The agent B is initially at a position $\vec x_B^0$, also at the side opposite to that from which the agent A was released, and off the path of A. It is brought to $\vec x_T$, at the position of the target, at an instant of time $t_B$ such that
\be
t_A^1<t_B<t_A^2 \, ,
\ee
and then brought back to $\vec x_B^0$.

For a configuration of the system satisfying the properties listed above, we can now determine the final state of the system at $t_f$. The initial state of the system at $t=0$ is
\be
\frac{1}{\sqrt{2}}\left( \ket{\mathcal{M}_1}+\ket{\mathcal{M}_2} \right) \ket{r_i}_A \ket{\vec x_B^0}_B \ket{\vec x_T}_T \ket{\psi}_T \, ,
\ee
where the state of A is labelled by the radial position of the agent, the state of B by its position, and the state of the target is the tensor product of a state describing its position and an internal state $\ket{\psi}_T$. The evolution in each branch of the geometry can be determined independently. In the first branch, the position of A follows the radial geodesic $\gamma_1$ in $\mathcal{M}_1$. In its first oscillation, after crossing the interior region, it meets the target at $t=t_A^1$, and applies the operation $\mathcal{A}$ on its internal state at this time. Next, it oscillates radially around the origin and returns to its initial position at $t=t_f$. The target system, after meeting the agent A, encounters the agent B at a later time $t_B>t_A^1$, when the agent applies the operation $\mathcal{B}$ at a proper time $\tau_B$. The final internal state of the target is then $\mathcal{B}\mathcal{A} \ket{\psi}_T$. The target and agent B are afterwards brought back to their initial positions. Hence, the evolution in this branch is given by:
\be
\ket{\mathcal{M}_1} \ket{r_i}_A \ket{\vec x_B^0}_B \ket{\vec x_T}_T \ket{\psi}_T  \mapsto \ket{\mathcal{M}_1} \ket{r_i}_A \ket{\vec x_B^0}_B \ket{\vec x_T}_T \mathcal{B}\mathcal{A} \ket{\psi}_T  \, .
\ee
In the second branch, the position of A follows the radial geodesic $\gamma_2$ in $\mathcal{M}_2$. It also meets the target system after crossing the interior region, but at a time $t_A^2>t_B$. In this branch the target first meets B and then A. Hence, its final internal state is $\mathcal{A}\mathcal{B} \ket{\psi}_T$. The agents and target return to their initial positions at $t_f$, after their interactions have taken place. The evolution in this branch is given by:
\be
\ket{\mathcal{M}_2} \ket{r_i}_A \ket{\vec x_B^0}_B \ket{\vec x_T}_T \ket{\psi}_T  \mapsto \ket{\mathcal{M}_2} \ket{r_i}_A \ket{\vec x_B^0}_B \ket{\vec x_T}_T \mathcal{A}\mathcal{B} \ket{\psi}_T  \, .
\ee

The final states of the agents and of the position of the target are the same in both branches, and disentangle from the rest of the system. The final state of the subsystem formed by the state of spacetime and the internal state of the target is
\be
\frac{ \ket{\mathcal{M}_1} \mathcal{B}\mathcal{A} \ket{\psi}_T +\ket{\mathcal{M}_2} \mathcal{A}\mathcal{B} \ket{\psi}_T }{\sqrt{2}} \, .
\ee
A quantum switch \eqref{eq:quantum-switch} is thus implemented, with the state of the geometry playing the role of the control bit. Assuming that the state of the geometry can be measured in a diagonal basis $\ket{\pm}=(\ket{\mathcal{M}_1} \pm \ket{\mathcal{M}_2})/\sqrt{2}$, the target system is then brought to the state
\be
\frac{\mathcal{B}\mathcal{A} \ket{\psi}_T \pm \mathcal{A}\mathcal{B} \ket{\psi}_T}{\sqrt{2}} \, ,
\ee
where the sign $\pm$ refers to the result of the measurement of the diagonal basis that was postselected. For either choice, the final internal state of the target is a superposition of those obtained with the application of the operations in switched orders. The superposition of orders can then be verified by performing observations of the internal state of the target system.

\begin{figure}
\begin{subfigure}{.5\textwidth}
  \includegraphics[width=.96\linewidth]{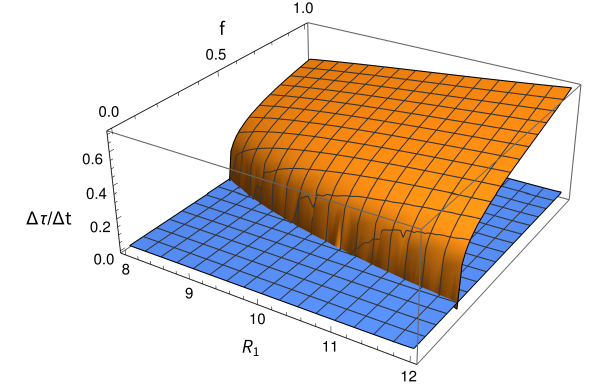}
  \label{fig:sub1}
\end{subfigure}%
\begin{subfigure}{.5\textwidth}
  \centering
  \includegraphics[width=.82\linewidth]{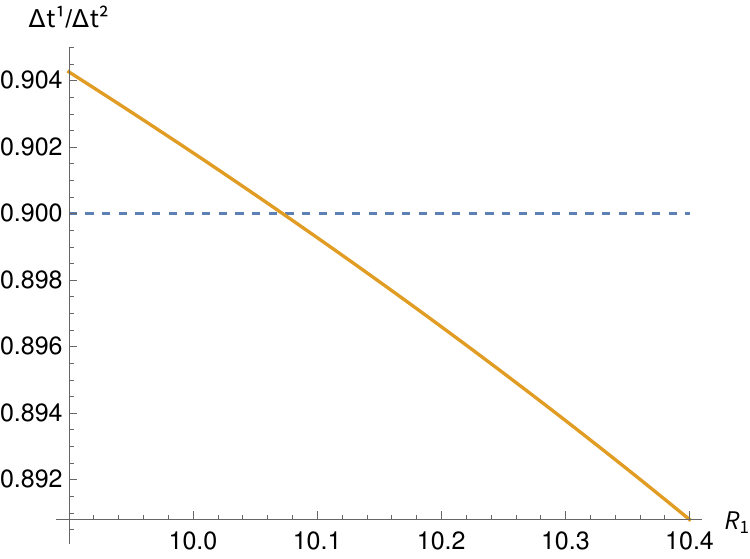}
  \label{fig:sub2}
\end{subfigure}
\caption{Left panel: Ratio between proper time and coordinate time elapsed in one oscillation on the spacetimes $\mathcal{M}_1$ (orange surface) and $\mathcal{M}_2$ (blue surface) in terms of the parameters $R_1,f$. The ratios are identical at the intersection of the two surfaces, which provides solutions to condition \eqref{eq:cond-1-a}. Right panel: Ratio of coordinate times elapsed in one oscillation on the spacetimes $\mathcal{M}_1$ and $\mathcal{M}_2$ in terms of $R_1$, with the parameter $f$ fixed so that condition \eqref{eq:cond-1-a} is satisfied. The ratio varies continuously, providing solutions to condition \eqref{eq:cond-1-b}.}
\label{fig:cond-1}
\end{figure}

Let us now show that geodesics satisfying the required properties exist. The parameters that can be varied are the masses $m$ and $M$ of the Schwarzschild patches, the radii $R,R_1,R_2$ of the shells and the initial radial position $r_i$ of the agent A. The first condition we impose is that after a number of oscillations, the geodesics $\gamma_1$ and $\gamma_2$ of the agent A in the spacetimes $\mathcal{M}_1$ and $\mathcal{M}_2$ meet again at the initial position in the exterior region, at the same coordinate and proper times. We do it in two steps. We first impose the condition
\be	\label{eq:cond-1-a}
\frac{\Delta \tau^{(1)}}{\Delta t^{(1)}}=\frac{\Delta \tau^{(2)}}{\Delta t^{(2)}}
\ee
that the ratio of proper time to coordinate time in one oscillation is the same in both geometries. This ensures that, if the geodesics return to the initial position at the same coordinate time, then they will also have the same proper time. In addition, we require that
\be	\label{eq:cond-1-b}
\frac{\Delta t^{(1)}}{\Delta t^{(2)}}= \frac{p}{q} \, , \qquad p,q \in \mathbb{N} \, ,
\ee
i.e., that the ratio of the periods of oscillation in the distinct geometries is a rational number. This ensures that the geodesics meet again at the initial position, at the same coordinate time, after $q$ oscillations in the spacetime $\mathcal{M}_1$ and $p$ oscillations in $\mathcal{M}_2$.

In order to numerically solve the conditions \eqref{eq:cond-1-a} and \eqref{eq:cond-1-b}, we proceeded as follows. We fixed the radius $R_2$ of the shell $\Sigma_2$, the masses $m,M$ and the initial radius $r_i$, and let the radii $R$ and $R_1$ of the shells $\Sigma$ and $\Sigma_1$ be free parameters that could be varied, allowing us to look for solutions for the conditions in this restricted parameter space. We further restricted to configurations satisfying $R_2 < R < R_1$ by writing
\be
R= R_2 + (R_1 - R_2)f \, , \qquad f \in [0,1] \, ,
\ee
and required that $R>2M$ and $R_2>2m$, so that the shells do not form a black hole in either geometry. We computed the periods of oscillation in coordinate and proper times on each spacetime, and plotted the ratios $\Delta \tau^{(1)}/\Delta t^{(1)}$ and $\Delta \tau^{(2)}/\Delta t^{(2)}$ in terms of the free parameters $R_1,f$. The intersection of these surfaces describe solutions of the condition \eqref{eq:cond-1-a}. Setting $m=1.9999$, $M=3$, $R_2=4$ and $r_i=12$, we obtained a family of solutions along a line on the plane $(R_1,f)$, as shown in the left panel of Figure \ref{fig:cond-1}. For these solutions, we can plot the ratio $\Delta t^{(1)}/\Delta t^{(2)}$ as a function of $R_1$, restricting to values such that $R>2M$. We verified that the ratio changes continuously, as shown in the right panel of Figure \ref{fig:cond-1}. Therefore, there is an infinite number of solutions with rational ratios $\Delta t^{(1)}/\Delta t^{(2)}$. As an example, a specific solution with $\Delta t^{(1)}/\Delta t^{(2)}=9/10$ is shown, for which $R_1=10.072$, $f=0.329464$ and $R=6.00057$.

\begin{figure}
\begin{subfigure}{.5\textwidth}
  \centering
  \includegraphics[width=.75\linewidth]{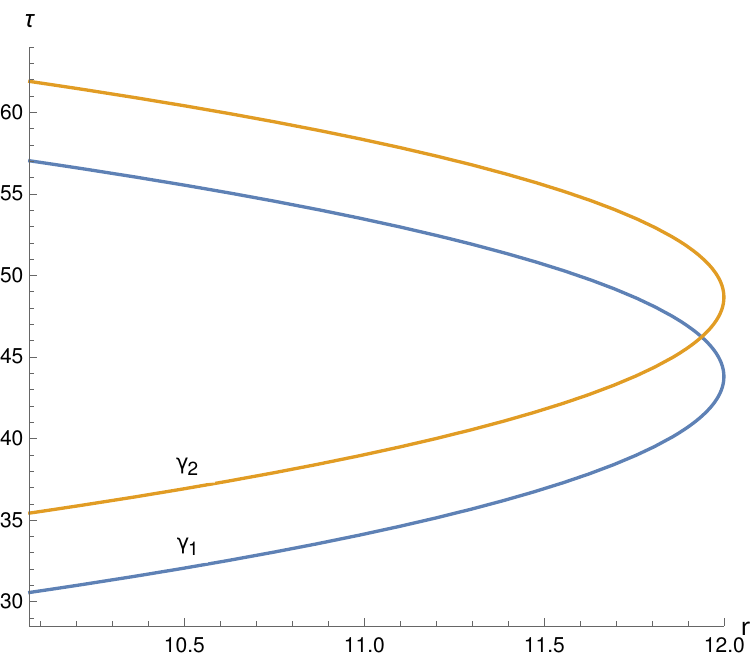}
  \label{fig:sub1}
\end{subfigure}%
\begin{subfigure}{.5\textwidth}
  \centering
  \includegraphics[width=.75\linewidth]{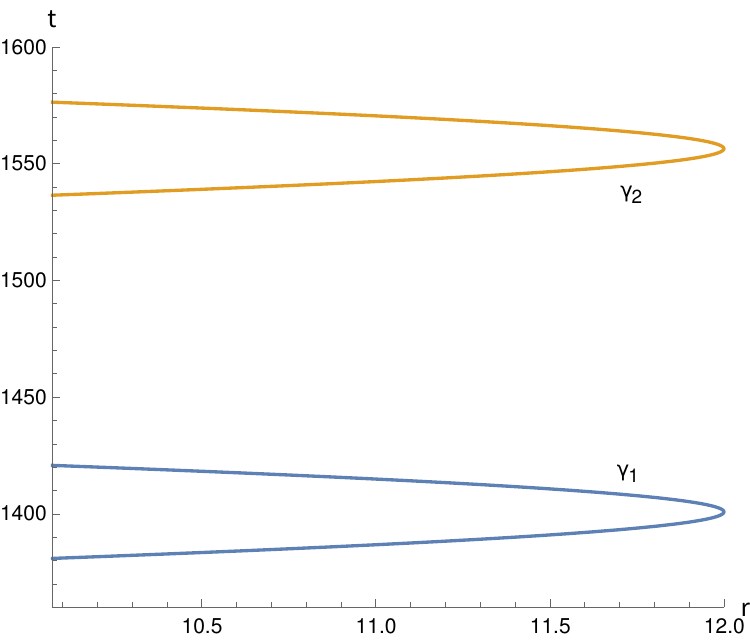}
  \label{fig:sub2}
\end{subfigure}
\caption{Proper time (left panel) and coordinate time (right panel) along the geodesics $\gamma_1$ and $\gamma_2$ on the spacetimes $\mathcal{M}_1$ and $\mathcal{M}_2$ in the exterior region $r>R_1$, after the first crossing of the interior region.}
\label{fig:cond-2}
\end{figure}

We also need to show that the geodesics $\gamma_1$ and $\gamma_2$ are such that, after first crossing the interior region $r<R_1$, both cross some radius $r_t$ at the same proper time $\tau_A$ but distinct coordinate times $t_A^1<t_A^2$, with $\gamma_1$ crossing $r_t$ while travelling back toward the interior region, and $\gamma_2$ crossing $r_t$ while travelling toward the turning point at $r_i$. We verified this explicitly for the solution with $\Delta t^{(1)}/\Delta t^{(2)}=9/10$ by plotting the coordinate and proper times as functions of the radius for each geodesic, as shown in Figure \ref{fig:cond-2}. In the left panel, the proper times are plotted against the radius. The geodesics intersect at the radius $r_t=11.9382$ in the desired manner. In the right panel, the coordinate times are plotted as functions of the radius. The whole trajectory of $\gamma_1$ in the exterior region takes place before $\gamma_2$ reaches the region, showing that $t_A^1<t_A^2$, as desired. This completes the verification of all properties required for the implementation of the quantum switch.

\subsection{Comparison with previous protocols}

A protocol for the implementation of a gravitational quantum switch was originally proposed in \cite{tbell}, where a superposition of geometries produced by a point mass in a superposition of two positions was considered. The transformation \eqref{eq:quantum-switch} that characterizes a general quantum switch has also been performed in nongravitational systems in recent experiments~\cite{RevExp}. In such experimental realizations, the process is implemented in optical tables, with distinct degrees of freedom of a photon, as for instance its path and polarization, used as the control bit and the target system. Distinct protocols for the implementation of the quantum switch are thoroughly discussed in \cite{Nikola}.

There are distinct ways of performing the task \eqref{eq:quantum-switch} in situations where a global time variable $t$ associated with an external observer is available, as in the case of optical table experiments, where $t$ is the local time in the laboratory where the experiment is realized. One can consider, for instance, a setup in which the operations $\mathcal{A}$ and $\mathcal{B}$ are each applied at two different times:
\begin{multline}
\label{QScircuitOpt}
\big(
\mathcal{A} \otimes \ket{0}\bra{0} + \mathds{1}_A \otimes\ket{1}\bra{1}
\big)|_{t=t_2} \big(
\mathds{1}_B\otimes\ket{0}\bra{0} + \mathcal{B} \otimes\ket{1}\bra{1}
\big)|_{t=t_2}
\\
\big(\mathds{1}_A \otimes \ket{0}\bra{0} + \mathcal{A} \otimes \ket{1}\bra{1}
\big)|_{t=t_1} \big(
\mathcal{B} \otimes\ket{0}\bra{0} + \mathds{1}_B\otimes\ket{1}\bra{1}
\big)|_{t=t_1}
\left(
  |\psi\rangle\otimes\frac{(|0\rangle+|1\rangle)}{\sqrt{2}}
\right) \, , 
\end{multline}
with the states $\ket{0},\ket{1}$ referring to the control bit. The two operations in the first line are applied at the same time $t_2$, and the two operations in the second line at the same time $t_1<t_2$. The composition of all operations implements the transformation \eqref{eq:quantum-switch}. The agents A and B are assumed to be at distinct positions, so that operations performed at equal times commute. The optical table implementations of the quantum switch reported in \cite{QS1,QS2} are realized in such a manner. Further discussions on such implementations and the ways of expressing them can be found in references~\cite{Nikola,OreshkovQuantum, NickAugustinJonathan, VilasiniRenner}.

Alternatively, one can consider a setup in which the operation $\mathcal{B}$ is applied at the same instant of time for both paths, and the operation $\mathcal{A}$ is applied at two distinct times:
\begin{multline}
\label{QScircuit}
\big(
\mathcal{A} \otimes \ket{0}\bra{0} + \mathds{1}_A \otimes\ket{1}\bra{1}
\big)|_{t=t_3}
\big(\mathcal{B} \otimes \ket{0}\bra{0} + \mathcal{B} \otimes\ket{1}\bra{1}\big)|_{t=t_2}
\\
\big(\mathds{1}_A\otimes\ket{0}\bra{0}+\mathcal{A} \otimes \ket{1}\bra{1} 
\big)|_{t=t_1}
\left(
  |\psi\rangle\otimes\frac{(|0\rangle+|1\rangle)}{\sqrt{2}}
\right) \, ,
\end{multline}
where $t_1<t_2<t_3$ refer to instants of time defined in the frame of an external observer. Examples of such an implementation are given by a gravitational quantum switch described in the Supplementary Material of reference~\cite{tbell}, as described in terms of global coordinates associated with an observer far away from the mass in superposition, and by the protocol proposed in \cite{MSY}. The protocol herein proposed, when described in terms of the reference frame associated with the local observers in the exterior region, also has this property.

A distinctive feature of a gravitational quantum switch based on an operational approach is that an event which is localized for one observer may appear as delocalized in time for another observer. Indeed, in both the protocol of ~\cite{tbell} and in our proposal, the operation $\mathcal{A}$ consists of a single event at a definite proper time for an observer in the laboratory A, while for an external observer, which in our case resides in the exterior region, such an event appears as delocalized in time. Ultimately, such a delocalization is what allows for the implementation of the superposition of orders, since in one branch of the superposition of geometries, the operation $\mathcal{A}$ is an event in the future of $\mathcal{B}$, and in the other branch of the superposition it is in the past of $\mathcal{B}$. That is, the switch is implemented with only two operationally defined events $\mathcal{A}$, $\mathcal{B}$, but one of the events appears in a superposition of two distinct times for an external observer.

A special property of our protocol is that the quantum switch can be implemented for arbitrary operations, even when controlled by the result of some other measurement performed in the laboratories. This results from the fact that each laboratory is acted by the same local gravity in both spacetimes in superposition, preventing the states of the geometry to be distinguished from within the laboratories. If the states of the geometry could be distinguished, they could be used to control the operation performed at the laboratiores, and a quantum switch would not be implemented. In order to discuss this point in more detail, let us consider a more general setup for the process described by Eq.~\eqref{QScircuit}. In the protocol \eqref{QScircuit}, the operations performed by A are controlled only by the coordinate time $t$, whose readings $t_3$ and $t_1$ in fact correspond to the same proper time $\tau^*$ of A for a gravitational quantum switch. A more general protocol can be considered for operations controlled by additional measurements: besides the proper time $\tau$, the agent can also measure some other degree of freedom $\lambda$, and apply operations according to the results obtained. In this case, we can consider a generalization of the form:
\begin{multline}
\label{QSgcircuit}
\big(
\mathcal{A}(\tau_1;\lambda_1)\otimes\ket{0}\bra{0} + \mathds{1}_A \otimes\ket{1}\bra{1}
\big)
\big(\mathcal{B}\otimes\ket{0}\bra{0} + \mathcal{B}\otimes\ket{1}\bra{1}\big)
\\
\big( \mathds{1}_A\otimes\ket{0}\bra{0} + \mathcal{A}(\tau_2;\lambda_2)\otimes\ket{1}\bra{1}
\big)
\left(
  |\psi\rangle\otimes\frac{(|0\rangle+|1\rangle)}{\sqrt{2}}
\right) \, ,
\end{multline}
where $\mathcal{A}(\tau;\lambda)$ means an operation performed at the proper time $\tau$ that depends on the degree of freedom $\lambda$. The protocol of~\cite{tbell} is a particular case of~\eqref{QSgcircuit}, with $\tau_1=\tau_2=\tau^*$ and  $\mathcal{A}(\tau_1;\lambda_1)=\mathcal{A}(\tau_2;\lambda_2)=\mathcal{A}$, so that $\mathcal{A}(\tau;\lambda)$ does not depend on $\lambda$ and is applied in both branches of the superposition at the same proper time $\tau=\tau^*$.

A slightly more general operation can be considered, however, with $\tau_1=\tau_2=\tau^*$, and $\mathcal{A}(\tau_1;\lambda_1)=\mathcal{A}(\lambda_1)$, $\mathcal{A}(\tau_2;\lambda_2)=\mathcal{A}(\lambda_2)$. In this case, the operation of A is still applied at the proper time $\tau=\tau^*$, but now it can depend on $\lambda$. Naturally, a quantum switch will not then be implemented if $\mathcal{A}(\lambda_1)\neq \mathcal{A}(\lambda_2)$. As an illustration, if the laboratory A chooses to perform an operation $\mathcal{A}(\lambda)$ such that $\mathcal{A}(\lambda_1)=\mathcal{C} \neq \mathcal{A}(\lambda_2)=\mathcal{D}$, then we obtain the following process:
\begin{equation}
\label{noQS}
\frac{\mathcal{C} \mathcal{B}|\psi\rangle\otimes|0\rangle+\mathcal{B} \mathcal{D}|\psi\rangle\otimes|1\rangle}{\sqrt{2}}
\end{equation}
which is not a quantum switch. Therefore, if we allow for arbitrary operations, the protocol does not necessarily implement a quantum switch. The parameter $\lambda$ can represent, in particular, a measurement of the weight of an object in the laboratory. Now, consider the protocol discussed in \cite{tbell}. There, the laboratories A and B move along wordlines with distinct accelerations, and the parameter $\lambda$ will assume distinct values for each geometry in superposition when representing a measurement of the local gravity. The superposition of orders will then be implemented only for a subset of all possible operations that the laboratories can perform, namely, those independent of $\lambda$. In this way, the implementation of the indefinite order depends not only on the structure of the spacetime, but also on the operations being applied. If the laboratory A is in free fall, on the other hand, the measurement of the local gravity has always the same outcome, and the same operation would be implemented in both branches of the superposition.

Note that the tabletop realizations of a quantum switch can also be described in terms of operations controlled by an additional degree of freedom $\lambda$, where $\lambda\in\{$yes, no$\}$ is a variable that describes if the photon is inside the laboratory or not, and $\tau=t$ is the global time. The operation applied by laboratory A can be represented as $\mathcal{A}(t;\lambda)=\mathcal{A}(\lambda)=\mathcal{A}\delta_{\lambda,\text{yes}}+\mathds{1}\delta_{\lambda,\text{no}}$, as it is performed only when the target system is found in the laboratory. The protocol can then be written as:
\begin{multline}
\big(
\mathcal{A}(t^*;\text{yes})\otimes\ket{0}\bra{0}+\mathcal{A}(t^*;\text{no}) \otimes\ket{1}\bra{1}
\big)
\big(\mathcal{B}\otimes\ket{0}\bra{0}+\mathcal{B}\otimes\ket{1}\bra{1}\big)
\\
\big(\mathcal{A}(t^{**};\text{no})\otimes\ket{0}\bra{0}+\mathcal{A}(t^{**};\text{yes})\otimes\ket{1}\bra{1}
\big)
\left(
  |\psi\rangle\otimes\frac{(|0\rangle+|1\rangle)}{\sqrt{2}}
\right) \, .
\end{multline}
Of course, if the agent A were to chose to apply a more general operation of the form $\mathcal{A}(t;\lambda)=\mathcal{A}(t)\delta_{\lambda,\text{yes}}+\mathds{1}\delta_{\lambda,\text{no}}$, with  $A(t^*) \neq A(t^{**})$, the quantum switch would not be implemented, and its realization again depends on the choice of operation performed in the laboratory A.

\section{Discussion}
\label{sec:discussion}

We introduced a new protocol for the implementation of a gravitational quantum switch. The protocol is formulated on a superposition of spacetimes $\mathcal{M}_1$ and $\mathcal{M}_2$ with geometries produced by distinct configurations of thin spherical mass shells. The geometries are isometric in an exterior region outside a radius $R_1$, where both are described by a Schwarzschild metric of mass $M$. Inside this radius, the geometries differ, so that $\mathcal{M}_1$ and $\mathcal{M}_2$ are not globally isometric. Both interior geometries are well behaved, including a flat core described by a Minkowski metric, surrounded by distinct patches of Schwarzschild metrics glued together. In the proposed protocol, an agent A initially in the exterior region first frelly falls from rest towards the masses. After crossing the interior region, it follows distinct geodesics in the exterior region for each spacetime in superposition. We showed that the configuration of the shells and the wordline of the target can be chosen, nonetheless, so that in both spacetimes the agent A meets the target system at the same proper time $\tau_A$, when it applies an operation $\mathcal{A}$ on it. Another agent B, travelling in the exterior region, also acts on the target system, applying an operation $\mathcal{B}$ on it. The geometries are such that the operation $\mathcal{A}$ is performed in the causal past of the operation $\mathcal{B}$ in the spacetime $\mathcal{M}_1$, and in the causal future of $\mathcal{B}$ in $\mathcal{M}_2$. In addition, at the end of the protocol, the agents naturally disentangle from the state of the target. The operations are thus applied on the target in reversed orders in the distinct branches of the superposition of geometries, leading to the implementation of a quantum switch controlled by the state of the geometry.

In a superposition of geometries, it is not immediately clear how to characterize a physical event. We adopted an operational approach for that purpose \cite{tbell,Rovelli}, as in related works on quantum gravity phenomenology. An event can then be characterized as an interaction between a laboratory and a target system at a definite proper time as measured in the laboratory. Accordingly, the operations $\mathcal{A}$ and $\mathcal{B}$ correspond to interactions taking place at specific proper times of the agents, understood as laboratories in the superposition of geometries, as in the previous protocol \cite{tbell}. In addition, we applied the operational approach to provide an explicit operational meaning to the coordinate system used to specify the paths of the agents and the target. This was possible in our setup due to the existence of isometric exterior regions in the spacetimes in superposition. A network of laboratories can be positioned in the exterior region. As each laboratory can be observed in either spacetime in superposition, an operationally defined event that is localized in both spacetimes identifies a pair of points $x^\mu_1 \in \mathcal{M}_1$ and $x^\mu_2 \in \mathcal{M}_2$ \cite{Giacomini1,Giacomini2}. The pair $(x^\mu_1,x^\mu_2)$ defines what is called the quantum coordinates of the event \cite{Hamette}. In our setup, the same coordinates are assigned to events in such laboratories in the exterior region in both spacetimes. This reduces the quantum coordinates to classical coordinates $(x^\mu_1=x^\mu,x^\mu_2=x^\mu)\mapsto x^\mu$ in this region. The meaning of these coordinates is then the same as in a classical geometry: if a laboratory with coordinates $x^\mu$ sends a light signal to another laboratory with coordinates $y^\mu$, the time delay between emission and absorption is the same as that for a lightlike geodesic connecting such points in the classical geometry of the exterior region.

In the proposed protocol,  an agent crosses the interior region that displays a superposition of geometries. A novel feature of the protocol is that the agent that experiences the superposition of geometries is in free fall. This allows for a more general class of operations $\mathcal{A},\mathcal{B}$ to be considered for the implementation of the quantum switch than in previous protocols. No information from the geometry can be obtained from within a laboratory in free fall, as the local gravity inside it vanishes, assuming that the laboratory is sufficiently small so that tidal forces can be neglected. This is true not only for a classical geometry, but also for a superposition of geometries, as no measurement performed within the laboratory can distinguish between the geometries in superposition. In contrast, if a laboratory were required to follow nongeodesic paths with distinct accelerations in the geometries in superposition, it could in principle measure the weight of an object, and apply an operation on the target controlled by the result of the weight measurement. A quantum switch would not be implemented in this case, as distinct operations would be realized in each branch of the superposition. Hence, a protocol for a quantum switch involving laboratories in nongeodesic motion in the superposition of geometries can be implemented only for operations that are insensitive to the local gravity. This restriction is not necessary for laboratories in geodesic motion. In this sense, the proposed protocol is universal, being applicable to arbitrary operations $\mathcal{A}$ and $\mathcal{B}$, even when controlled by measurements performed in the laboratory. From a conceptual side, a laboratory in free fall is a natural realization of the concept of closed laboratory as explored in the process matrix formalism \cite{Oreshkov}, and our protocol corresponds to an implementation of the quantum switch as represented in such a formalism.

The interactions among the agents and the target system take place in the exterior region of the spacetimes, after the agent A has crossed the interior region that displays a superposition of geometries. The fact that the operations are applied in a region of spacetime with a definite geometry may seem to imply that the superposition of orders is unrelated to the superposition of geometries. That this is not the case, however, can be seen as follows. In a classical spacetime, a given event being in the causal past or future of another event is not a property of the metric at the events, but a global property of the spacetime. In a superposition of spacetimes, the superposition of orders results from that of such global structures, and not from fluctuations of the lightcone at the moment when the operations are performed. In particular, the temporal relation between two operationally defined events $\mathcal{A}$ and $\mathcal{B}$ in a pair of laboratories depends on the wordlines followed by the laboratories in each geometry in superposition, which cross an extended region of spacetime. In short, a superposition of orders between operationally defined events, with a quantum geometry used as a control bit, is not determined by a local superposition of lightcones at the events where the operations are applied, but by the superposition of the global structures of the spacetimes.

An eventual experimental realization of the proposed protocol would require the preparation of a large mass in a quantum superposition of two configurations in which the mass is distributed over a single or two spherical shells, which is beyond present-day experimental capabilities. Nonetheless, there is no fundamental reason preventing this to be achievable with future quantum technology. Some barriers currently being challenged for the production of macroscopic states of quantum matter are the production of superposition states for systems with large masses \cite{Yaakov} and superpositions of positions separated by large distances \cite{Kovachy}. Geometrical traps of complex shapes have also been engineered for cold atoms and Bose-Einstein condensates \cite{Henderson}, including a potential that traps atoms on a single spherical shell, which allowed the observation of bubbles of ultracold atoms on a microgravity environment \cite{Carollo}.

\acknowledgments

N.S.M. thanks M. Hamed Mohammady for useful suggestions, and acknowledges financial support from Štefan Schwarz Support Fund, project DeQHOST APVV-22-0570, project DESCOM VEGA 2/0183/21, project QuaSiModo VEGA 2/0156/22, and Grant No. 61466 from the John Templeton Foundation, as part of the “The Quantum Information Structure of Spacetime (QISS)” project (qiss.fr). The opinions expressed in this publication are those of the author(s) and do not necessarily reflect the views of the John Templeton Foundation.
B.S. acknowledges financial support from Coordena\c{c}\~ao de Aperfei\c{c}oamento de Pessoal de N\'{i}vel Superior (CAPES), Programa de Excelência Acadêmica (PROEX), under the Process No.~88887.495426/2020-00.
N.Y. acknowledges financial support from the Conselho Nacional de Desenvolvimento Cient\'ifico e Tecnol\'ogico (CNPq) under Grant No.~306744/2018-0.

\bibliographystyle{quantum}

\appendix

\section{Energy-momentum tensor of the shells}
\label{sec:energy-momentum}

Consider the spacetime $\mathcal{M}_1$. At the radius $r=R$, the extrinsic curvature $K^\pm_{ab}$ can be directly computed at each side of the surface $\Sigma$ from the definition \eqref{eq:K-pm}. We find that it is discontinous, with a discontinuity
\be	\label{eq:[K]-sigma}
[K_{ab}] = \diag\left[ -\frac{M}{R^2} \sqrt{1-\frac{2M}{R}}, R \left( \sqrt{1-\frac{2M}{R}} -1 \right), R \sin^2 \theta \left( \sqrt{1-\frac{2M}{R}} -1 \right) \right] \, .
\ee
The energy-momentum tensor expressed in the coordinates $x^\alpha=(t,\ell, \theta,\varphi)$ is given by Eqs.~\eqref{eq:mass-shell-energy-mom}, \eqref{eq:S-def} and \eqref{eq:[K]}. Its nonzero components read:
\begin{align}	\label{eq:T-sigma}
T^{00} &= \delta(\ell) \frac{1}{4\pi R} \left( 1-\sqrt{1-\frac{2M}{R}} \right) \left( 1-\frac{2M}{R}\right)^{-1} \, , \nonumber \\
T^{22} &= \delta(\ell) \frac{1}{8\pi R^3} \left( 1-\frac{M}{R}-\sqrt{1-\frac{2M}{R}} \right) \left( 1-\frac{2M}{R}\right)^{-1/2} \, , \nonumber \\
T^{33} &= \frac{1}{\sin^2 \theta} T^{22} \, .
\end{align}

The energy density $\rho$ and pressure $P$ in the shell can be computed from Eq.~\eqref{eq:T-sigma}. The $4$-velocity of an observer standing at the shell is given by
\[
u_\alpha=\diag\left[ -\left( 1-\frac{2M}{R}\right)^{1/2},0,0,0 \right] \, ,
\]
from which we obtain
\be
\rho = T^{\alpha \beta} u_\alpha u_\beta = \delta(\ell) \frac{1}{4 \pi R} \left( 1-\sqrt{1-\frac{2M}{R}} \right) \, .
\ee
For large radii $R \gg 2M$, the energy density reduces to $\delta(\ell)M/(4\pi R^2)$, as could be expected. Taking unit vectors $e_r,e_\theta,e_\varphi$ along the spatial directions, we can also determine the pressure in each direction. For the directions tangential to $\Sigma$, we find:
\begin{align}
P &= T^{\alpha \beta} (e_\theta)_\alpha (e_\theta)_\beta = T^{\alpha \beta} (e_\varphi)_\alpha (e_\varphi)_\beta \nonumber \\
	&= \delta(\ell) \frac{1}{8 \pi R} \left( 1-\frac{2M}{R}\right)^{-1/2} \left( 1-\frac{M}{R}-\sqrt{1-\frac{2M}{R}} \right) \, .
\end{align}
The tangential pressure $P$ vanishes at large radii and diverges at the horizon. The pressure vanishes in the orthogonal radial direction.

In the spacetime $\mathcal{M}_2$, the second junction condition is not satisfied at $\Sigma_1$ or $\Sigma_2$. The case of $\Sigma_2$ is identical to the previous case: a Schwarzschild patch is glued around an interior spherical region in Minkowski spacetime. The formulas for the discontinuity of the extrinsic curvature and the energy-momentum tensor can then be obtained from Eq.~\eqref{eq:[K]-sigma} and \eqref{eq:T-sigma} with the substitutions $R \to R_2$ and $M\to m$. For the surface $\Sigma_1$, the extrinsic curvature on each side of the surface read:
\begin{align*}
K^1_{ab} &= \sqrt{1-\frac{2M}{R_1}} \diag \left[ -\frac{M}{R_1^2}, R_1, R_1 \sin^2 \theta \right] \, , \\
K^2_{ab} &= \sqrt{1-\frac{2m}{R_1}} \diag \left[ -\frac{m}{R_1^2} \left( 1-\frac{2M}{R_1} \right) \left( 1-\frac{2m}{R_1} \right)^{-1}, R_1, R_1 \sin^2 \theta \right] \, .
\end{align*}
Therefore, $K^1_{ab}-K^2_{ab} \neq 0$, and a thin shell is present at $\Sigma_1$. The nonzero components of the energy-momentum tensor on $\Sigma_1$ are given by:
\begin{align}
T^{00}_1 &= \delta(\ell) \frac{1}{4\pi R_1} \left( \sqrt{1-\frac{2m}{R_1}}-\sqrt{1-\frac{2M}{R_1}} \right)  \left( 1-\frac{2M}{R_1}\right)^{-1} \, , \nonumber \\
T^{22}_1 &= \delta(\ell) \frac{1}{8 \pi R_1^4} \left( \frac{R_1-M}{\sqrt{1-\frac{2M}{R_1}}} - \frac{R_1-m}{\sqrt{1-\frac{2m}{R_1}}} \right) \, , \nonumber \\
T^{33}_1 &= \frac{1}{\sin^2 \theta} T^{22}_1 \, ,
\end{align}
where we used coordinates $x^\alpha=(t,\ell,\theta,\varphi)$ in a neighborhood of $\Sigma_1$, with
\begin{equation}
t=
\begin{cases}
t_1 \, , & \text{in region I} \, ,
\\
\left(1-\frac{2M}{R_1}\right)^{-1/2}\left(1-\frac{2m}{R_1}\right)^{1/2}t_2 \, , & \text{in region II} \, .
\end{cases}
\end{equation}
and where $\ell$ is the proper length along the radius, defined by
\[
\frac{d\ell}{dr_1} = \left(1-\frac{2M}{r_1}\right)^{-1/2} \, , \quad \frac{d\ell}{dr_2} = \left(1-\frac{2m}{r_2}\right)^{-1/2} \, ,
\]
and the boundary condition that $\ell=R_1$ at $\Sigma_1$. The metric is continuous in these coordinates.

The energy density $\rho_1$ and tangential pressure $P_1$ can be determined as before. We find that
\be
\rho_1 = \delta(\ell) \frac{1}{4\pi R_1} \left( \sqrt{1-\frac{2m}{R_1}}-\sqrt{1-\frac{2M}{R_1}} \right) \, .
\ee
For large radii $R_1 \gg 2M > 2m$, the energy density reduces to $\delta(\ell)(M-m)/(4\pi R_1^2)$. The radial component of the pressure, orthogonal to $\Sigma_1$, vanishes. The pressure in any tangential direction is given by
\be
P_1 = \delta(\ell) \frac{1}{8 \pi R_1^2} \left( \frac{R_1-M}{\sqrt{1-\frac{2M}{R_1}}} - \frac{R_1-m}{\sqrt{1-\frac{2m}{R_1}}} \right) \, .
\ee
It vanishes for $R_1 \gg 2M > 2m$, and diverges for $R_1= 2M$.

\end{document}